\documentclass[a4paper,11pt]{article}
\usepackage{jheppub} 
\usepackage[T1]{fontenc} 
\usepackage{multirow}
\usepackage{color}
\usepackage{ulem}
\usepackage{rotating}
\usepackage{booktabs}
\usepackage{dcolumn}
\usepackage{array} 
\usepackage{multirow}
\usepackage{lineno} 
\usepackage{siunitx}
\usepackage{amsmath}
\usepackage{bm}
\usepackage{mathrsfs}
\usepackage{graphicx}
\usepackage{comment}
\usepackage{float}
\DeclareSIUnit{\bmm}{\bm{m}}

\DeclareSIUnit{\clight}{\textnormal{\textit{c}}}

\newcommand{\BESIIIorcid}[1]{\href{https://orcid.org/#1}{\hspace*{0.1em}\raisebox{-0.45ex}{\includegraphics[width=1em]{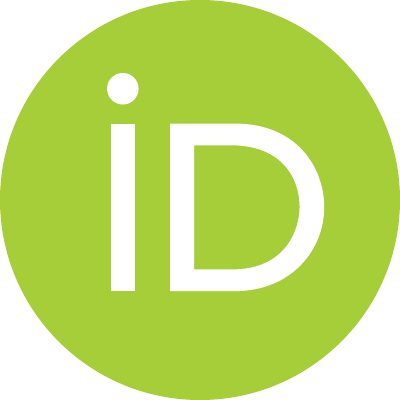}}}}
\newcolumntype{d}{D{.}{.}{-1}}
\newcolumntype{e}{D{.}{.}{8}}
\newcolumntype{f}{D{.}{.}{18}}
\newcolumntype{h}{D{.}{.}{13}}
\newcolumntype{g}{D{.}{.}{12}}
\title{\protect\boldmath Measurement of the cross sections of $\EE\ar K_{S}^{0}\bar\Xi^{0}\Lambda/\Sigma^{0} + \text{c.c.}$ at center-of-mass energies between $\mathbf{3.510}$ and \si{\mathbf{4.951}\,{\textbf{GeV}}}}

\collaborationImg{\includegraphics[width=0.25\textwidth]{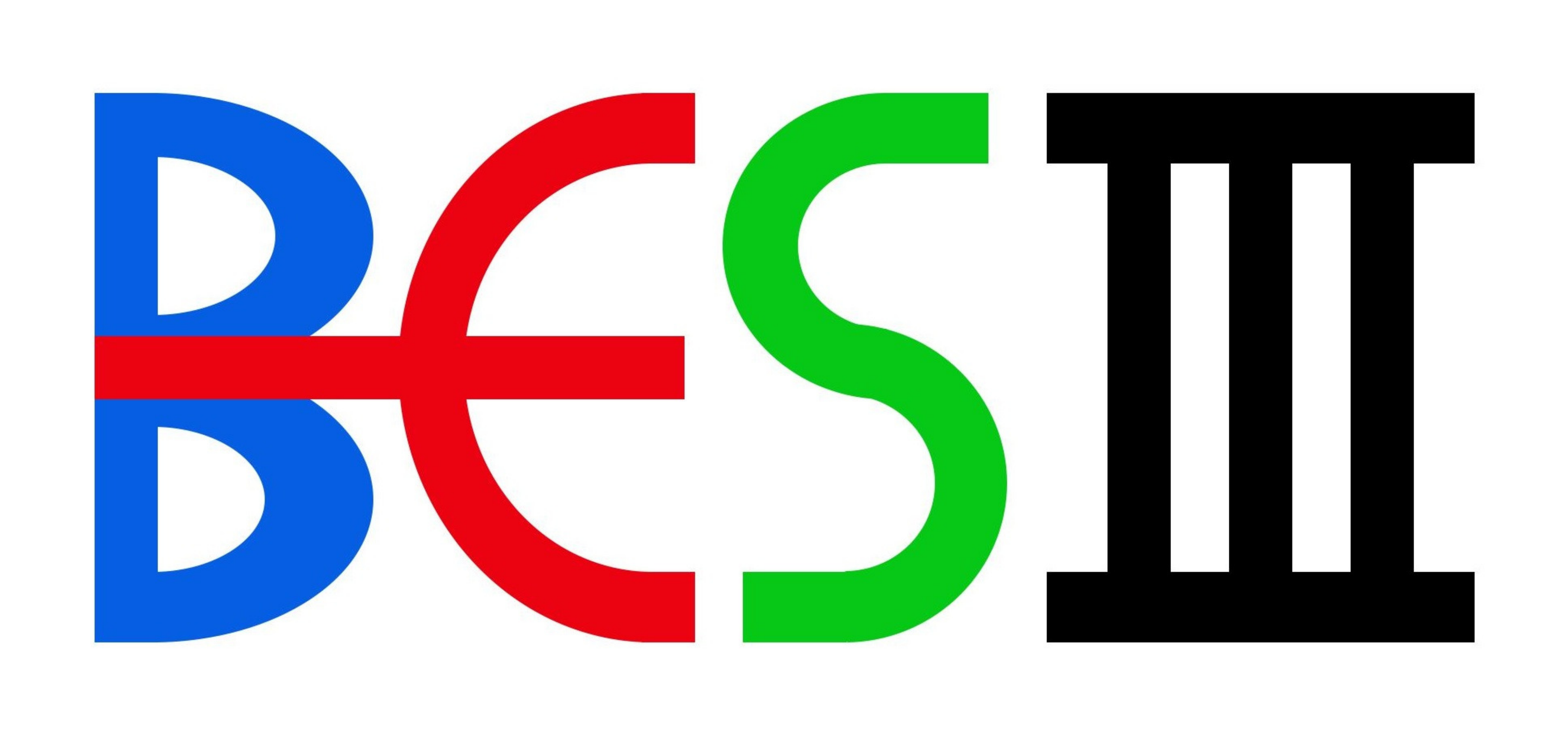}}
\collaboration{The BESIII Collaboration}

\emailAdd{besiii-publications@ihep.ac.cn}

\begin{document} 
\abstract{
Using $\EE$ collision data samples collected with the BESIII detector at the BEPCII at center-of-mass energies between 3.510 and \SI{4.951}{GeV} corresponding to an integrated luminosity of \SI{44.55}{fb^{-1}}, the Born cross sections of the processes \(e^+e^- \to K_S^0 \bar{\Xi}^0 \Lambda/\Sigma^0+\text{c.c.} \) are measured with a partial-reconstruction strategy. The dressed cross sections for the channels \(e^+e^- \to K_S^0 \bar{\Xi}^0 \Lambda/\Sigma^0 + \text{c.c.}\) are fitted with a model consisting of a power-law function and a charmonium (-like) resonance, considering the candidates $\psi(3770)$, $\psi(4040)$, $\psi(4160)$, $Y(4230)$, $Y(4360)$, $\psi(4415)$, $Y(4500)$, $Y(4660)$, and $Y(4710)$. No significant resonance contribution is observed in any of the fits. The upper limits for the products of the electronic partial widths and branching fractions at the 90\% confidence level are provided.
}

\clearpage{}

\newcommand{\Xib}{\bar\Xi^{0}}
\newcommand{\KXL}{\EE\ar K_{S}^{0}\bar\Xi^{0}\Lambda}
\newcommand{\KXS}{\EE\ar K_{S}^{0}\bar\Xi^{0}\Sigma^{0}}
\newcommand{\KXX}{\EE\ar K_{S}^{0}\bar\Xi^{0}\Lambda/\Sigma^{0}}
\newcommand{\EE}{e^+e^-}
\newcommand{\ar}{\rightarrow}
\newcommand{\bbt}{\bibitem}
\newcommand{\KS}{K_{S}^{0}}

\clearpage{}
\maketitle
\flushbottom
\section{Introduction}
\label{sec:intro}
The study of charmonium(-like) states produced in $\EE$ annihilation and decaying into baryonic final states provides important input for testing Quantum Chromodynamics and offers a valuable opportunity to investigate the nature of $XY\!Z$ states~\cite{Brambilla:2010cs, Briceno:2015rlt, XYZ:states, PR_2020, Recent_XYZ, XYZ_RMP98}. 
The potential model~\cite{Barnes:2005pb} predicts five vector charmonium states in the mass region between \SI{3.773}{GeV/\clight^{2}} and \SI{4.700}{GeV/\clight^{2}}, corresponding to the $3S$, $2D$, $4S$, $3D$, and $5S$ excitations~\cite{Brambilla:2010cs}. However, more vector states have been observed experimentally in this energy region than predicted by the potential model, suggesting the possible existence of exotic states. The three conventional charmonium states, $\psi(4040)$, $\psi(4160)$ and $\psi(4415)$, have been observed in measurements of the hadronic cross sections~\cite{BES:2001ckj,RVUE:2005,RVUE:2010, BES:2008, D0D0_2024}. 
Five unconventional charmonium-like states, $Y(4230)$, $Y(4360)$, $Y(4500)$, $Y(4660)$ and $Y(4710)$, have been observed in hidden-charm final states, either via initial-state radiation (ISR) at BaBar and Belle~\cite{BaBar:2005hhc, Belle:2007dxy, BaBar:2006ait, Belle:2007umv, BaBar:2012hpr, Belle:2013yex, Belle:2014wyt, BaBar:2012vyb}, or through direct production at CLEO~\cite{CLEO} and BESIII~\cite{BESIIIAB, BESIII:cpc1,BESIII:2023cmv,BESIII:2023cmv1}. 
These newly observed states cannot be interpreted as conventional charmonium resonances composed purely of a $c\bar{c}$ quark pair. To explain their nature, various interpretations have been proposed, including hybrid states, multiquark states, and hadronic molecules~\cite{Brambilla:2010cs,Briceno:2015rlt,Chen:2016qju,Wang:2019mhs,Close:2005iz,Qian:2021neg}. 
However, no definite conclusion has been reached so far, and the underlying nature of these states remains unclear. 
This situation reflects the limited understanding of the strong interaction in the non-perturbative regime. More high-precision experimental measurements are therefore needed to advance our knowledge. 
Among these studies, baryonic decays of charmonium (-like) states, dominated by three-gluon or one-photon annihilation processes, are particularly interesting due to the simple topologies of their final states compared to three-meson production. 
Although many studies of baryonic final states have been performed at BESIII~\cite{BESIII:2021ccp, Ablikim:2019kkp, Ablikim:2013pgf, BESIII:2017kqg, Wang:2021lfq, Wang:2022bzl, zhang:2026ssb, BESIII:2023rse, BESIII:2024umc, BESIII:2022kzc, Wang:2022zyc, BESIII:kxls, BESIII:2024sigma0, BESIII:2025ruoyu, BESIII:2025hl}, only a few decay modes have been firmly observed, such as $\psi(3770)\to\Sigma^-\bar{\Sigma}^+$~\cite{zhang:2026ssb}. 
In addition, evidence has been reported for the processes $\psi(3770)\to\Lambda\bar\Lambda$~\cite{BESIII:2021ccp}, $\psi(3770)\to\Xi^-\bar\Xi^+$~\cite{BESIII:2023rse}, and $\psi(4160)\to K^- \bar{\Xi}^+ \Lambda$~\cite{BESIII:kxls}, while no significant baryonic decays have been found for other vector charmonium(-like) states.
Therefore, precise measurements of the cross sections for exclusive baryonic final states in $e^{+}e^{-}$ annihilation above the open-charm threshold are crucial, as they provide valuable information for understanding the nature of vector charmonium(-like) states.

In this article, the Born cross sections of the processes $\KXL$ and $\KXS$ (charge-conjugate processes are implied throughout) are measured using $e^+e^-$ collision data corresponding to an integrated luminosity of \SI{44.55}{fb^{-1}}~\cite{ene1, BESIII:2022dxl, ene3,ene4,ene5}, collected with the BESIII detector~\cite{besiii} at the BEPCII collider~\cite{BEPCII} at center-of-mass (CM) energies $\sqrt{s}$ between 3.510 and \SI{4.951}{GeV}~\cite{ene1, ene3}. In addition, searches for vector resonances are performed by fitting the dressed cross sections of the $\KXX$ processes. Upper limits at the 90\% confidence level (C.L.) are determined for the products of the electronic partial widths and branching fractions of charmonium(-like) states decaying into the $\KS \Xib \Lambda/\Sigma^0$ final states.

\section{BESIII detector and Monte Carlo simulation}
The BESIII detector~\cite{besiii} records symmetric $e^+e^-$ collisions provided by the BEPCII storage ring~\cite{BEPCII} in the CM energy range from 1.84 to \SI{4.95}{GeV}, with a peak luminosity of \SI{1.1e33}{\per\centi\meter\squared\per\second} achieved at $\sqrt{s} =$ \SI{3.773}{GeV}. 
Large data samples have been collected in this energy region~\cite{Ablikim:2019hff, EcmsMea, EventFilter}. 
The cylindrical core of the BESIII detector covers 93\% of the full solid angle and consists of a helium-based multilayer drift chamber~(MDC), a time-of-flight system~(TOF), and a CsI(Tl) electromagnetic calorimeter~(EMC), which are all enclosed in a superconducting solenoidal magnet providing a 1.0~T magnetic field.
The solenoid is surrounded by an octagonal flux-return yoke made of steel, interleaved with resistive-plate-counter muon-identification modules.
The charged-particle momentum resolution at $1~{\rm GeV}/c$ is $0.5\%$, and the ${\rm d}E/{\rm d}x$ resolution is $6\%$ for electrons from Bhabha scattering. The EMC measures photon energies with a resolution of $2.5\%$ ($5\%$) at $1$~GeV in the barrel (end cap) region. 
The time resolution in the plastic scintillator TOF barrel region is 68~ps, while that in the end cap region was 110~ps.
The end cap TOF system was upgraded in 2015 using multigap resistive plate chamber technology, providing a time resolution of 60~ps, which benefits 81.5\% of the data used in this analysis~\cite{etof3}.

Monte Carlo (MC) simulated data samples produced with a {\sc geant4}-based~\cite{GEANT4} software package, which includes the geometric description of the BESIII detector~\cite{Huang:2022wuo} and the detector response, are used to determine detection efficiencies and to estimate backgrounds. The simulation models the beam energy spread and ISR in the $e^+e^-$ annihilations with the generator {\sc kkmc}~\cite{KKMC}.
The inclusive MC sample includes the production of $D\bar{D}$ pairs (including quantum coherence for the neutral $D$ channels), the non-$D\bar{D}$ decays of the $\psi(3770)$, the ISR production of the $J/\psi$ and $\psi(3686)$ states, and the continuum processes incorporated in {\sc kkmc}~\cite{KKMC}.
The detection efficiencies of the reactions $\KXX$ are determined by MC simulations. A sample of $200,000$ signal events is simulated with a phase-space (PHSP) distribution for each energy point, where the $\bar\Xi^{0}$ baryon, $\bar{\Lambda}$ baryon (from $\bar\Xi^{0}$) and $K_S^0$ meson with their subsequent decays to $\bar\Lambda \pi^0$, $\bar{p}\pi^+$ and $\pi^+\pi^-$ are described by the {\sc evtgen} program~\cite{evtgen2,EVTGEN} with a PHSP model.

\section{Event selection}
A partial-reconstruction technique is employed to select the $\KXX$ candidate events, where the $\Xib$ baryon is reconstructed from its $\bar\Lambda\pi^0$ decay mode with the subsequent $\bar\Lambda\to \bar p\pi^+$ decay and the $K_S^0$ meson is reconstructed from its $\pi^+\pi^-$ decay mode. The presence of the $\Lambda/\Sigma^{0}$ baryon is inferred from the invariant mass of the system recoiling against the reconstructed $K_{S}^0\Xib$ system. 

The charged tracks detected in the MDC are required to be within a polar angle ($\theta$) range of $|\!\cos\theta| < 0.93$, where $\theta$ is defined with respect to the $z~$axis, which is the symmetry axis of the MDC. 
Since a partial reconstruction technique is employed, at least two positively and two negatively charged tracks are required. 

Particle identification~(PID) for charged tracks combines measurements of the energy deposited in the MDC~(d$E$/d$x$) and the flight time in the TOF to form likelihoods $\mathcal{L}(h)~(h=p,K,\pi)$ for each hadron $h$ hypothesis.
Tracks are identified as protons when the proton hypothesis has the greatest likelihood ($\mathcal{L}(p)>\mathcal{L}(K)$ and $\mathcal{L}(p)>\mathcal{L}(\pi)$), and charged pions are identified by requiring $\mathcal{L}(\pi)>\mathcal{L}(p)$ and $\mathcal{L}(\pi)>\mathcal{L}(K)$.
Events with at least one $\bar{p}$ candidate, one $\pi^-$ candidate, and two $\pi^+$ candidates are retained for further analysis.

Photon candidates are identified using isolated showers in the EMC. The deposited energy of each shower must be more than 25~MeV in the barrel region ($|\cos \theta|< 0.80$) and more than 50~MeV in the end cap region ($0.86 <|\cos \theta|< 0.92$). To suppress electronic noise and showers unrelated to the event, the difference between the EMC time and the event start time is required to be within $[0, 700]\,\text{ns}$. After applying these selections, at least two photons are required.

The $\pi^0$ candidates are reconstructed via a one-constraint (1C) kinematic fit by looping over all pairs of photon candidates and constraining $M_{\gamma\gamma}$ to $m_{\pi^0}$, where $M_{\gamma\gamma}$ is the invariant mass of the $\gamma\gamma$ combination and $m_{\pi^0}$ is the nominal mass of $\pi^0$ meson taken from the Particle Data Group (PDG)~\cite{PDG2020}. All $\gamma\gamma$ combinations with a successful fit ($\chi^2_{\mathrm{1C}} < 20$~\cite{BESIII:2016nix,BESIII:2019dve,BESIII:2021aer}) are chosen as the $\pi^0$ candidates.

Candidates for the $\bar \Lambda$ baryon and the $K_S^0$ meson are formed by combining two oppositely charged tracks into the final states $\bar{p}\pi^+$ and $\pi^+\pi^-$, respectively. All $\bar{p}\pi^+$ and $\pi^+\pi^-$ combinations are used to reconstruct $\bar{\Lambda}$ and $K_S^0$ via secondary-vertex fits~\cite{vtxfit}, respectively. Combinations in which both candidates share the same $\pi^+$ track are discarded. If multiple candidates are present in an event, the best candidate is selected by minimizing the sum of the $\chi^2$ values from the two secondary-vertex fits. 
To suppress backgrounds from non-$\bar{\Lambda}$ and non-$K_S^0$ sources, the decay length of the $\bar{\Lambda}$ and $K_S^0$ candidates must be greater than twice the vertex resolution. Here, the decay length is measured as the distance between the production and decay vertices obtained from the secondary-vertex fit. The invariant mass of the $\bar{p}\pi^+$ pair is required to be within \SI{\pm5}{MeV/\clight^{2}} of the nominal mass of the $\bar{\Lambda}$ baryon, while that of the $\pi^+\pi^-$ pair is required to be within \SI{\pm10}{MeV/\clight^{2}} of the nominal mass of the $K_S^0$ meson, with nominal values taken from the PDG~\cite{PDG2020}. The mass windows correspond to approximately three times the mass resolutions determined from signal MC simulations.

To reconstruct the $\bar{\Xi}^{0}$ candidate, all $\bar{\Lambda}\pi^{0}$ combinations are examined, and the one that minimizes $\left| M_{\bar{\Lambda}\pi^{0}} - m_{\bar{\Xi}^{0}} \right|$ is selected, where $M_{\bar{\Lambda}\pi^{0}}$ is the invariant mass of the $\bar{\Lambda}\pi^{0}$ combination and $m_{\bar{\Xi}^{0}}$ is the nominal mass of the $\bar{\Xi}^{0}$ baryon taken from the PDG~\cite{PDG2020}. Moreover, the $\Xib$ signal region in the $M_{\bar{\Lambda}\pi^0}$ distribution is defined within \SI{\pm15}{MeV/\clight^{2}} of $m_{\bar{\Xi}^{0}}$, corresponding to about three times the mass resolution obtained from signal MC simulations. The $\Lambda/\Sigma^{0}$ candidates are inferred from the mass recoiling against the $K_S^{0}\bar{\Xi}^{0}$ system, defined as
\begin{linenomath*}
\begin{equation}
M^{\mathrm{Recoil}}_{K_S^{0}\bar{\Xi}^{0}}
 = \sqrt{\left(\sqrt{s}-E_{K_S^{0}\bar{\Xi}^{0}}\right)^{2}
 - \left|\vec{p}_{K_S^{0}\bar{\Xi}^{0}}\right|^{2}},
\end{equation}
\end{linenomath*}
where $E_{K_S^{0}\bar{\Xi}^{0}}$ and $\vec{p}_{K_S^{0}\bar{\Xi}^{0}}$ are the energy and momentum of the selected $K_S^{0}\bar{\Xi}^{0}$ system in the $e^{+}e^{-}$ CM frame, respectively.

After applying all event selection criteria, the remaining backgrounds predominantly originate from processes with final-state topologies similar to that of the signal channel, such as $K_S^0\bar{\Delta}^{0}\Sigma^{0}$, $\gamma\psi(2S)$, and $\pi^{+}K_S^{0}\bar{p}\Lambda$. However, these background contributions are found to be smoothly distributed in the signal region of the $M^{\mathrm{Recoil}}_{K_S^{0}\bar{\Xi}^{0}}$ spectrum.

\section{Born cross-section measurement}
\subsection{Extraction of signal yields}
\begin{figure}[!htbp]
  \begin{center}
  \includegraphics[width=0.80\textwidth]{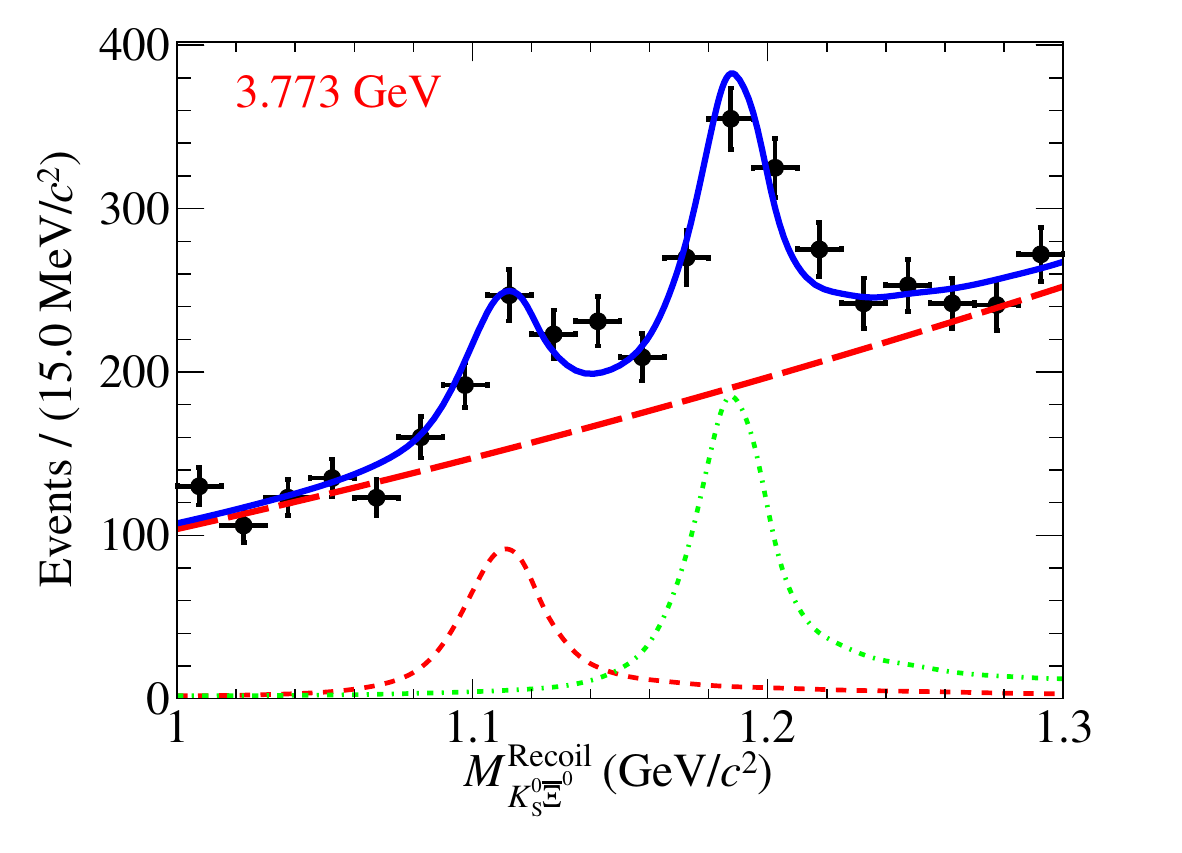} 
  \vspace*{-0.5cm}
  \end{center}
  \caption{
Fit to the $M^{\rm Recoil}_{K^0_S\bar\Xi^0}$ distribution from data at $\sqrt{s} = 3.773~\mathrm{GeV}$, where the black dots with error bars are data, the blue line represents the fitting results, the red short dashed line denotes the process $e^+e^- \rightarrow K_{S}^0 \bar{\Xi}^0 \Lambda$, the green dash-dotted line denotes the process $e^+e^- \rightarrow K_{S}^0 \bar{\Xi}^0 \Sigma^0$, and the red long dashed line represents the background contribution.}
  \label{Fig:SP:DATA:fitting}
  \end{figure}

To obtain the $\EE\ar K_S^0\bar{\Xi}^0\Lambda/\Sigma^{0}$ signal yield, an unbinned maximum likelihood fit is performed to the $M^{\rm Recoil}_{K_S^0\Xib}$ spectrum in the range from 1.0 to \SI{1.3}{GeV/\clight^{2}}.
In the fit, the signal shapes for the channels $\KXX$ at each energy point are described by the MC simulated shapes, and a second-order polynomial function describes the background shape. As an example, figure~\ref{Fig:SP:DATA:fitting} shows the fit result for the extraction of signal yields at $\sqrt{s} = \SI{3.773}{GeV}$, and 
the signal yields at all 56 energy points are listed in table~\ref{tab:NUM_BCSL} and table~\ref{tab:NUM_BCSS}. 
The data points near the $\psi(3686)$ resonance are excluded from this analysis since the energy points below the $\psi(3770)$ are primarily used to constrain the energy dependence of the continuum production. 
The statistical significance of the signal is evaluated by comparing the likelihood values of fits performed with and without the signal component, considering the change in the number of degrees of freedom. As the statistical significance at most energy points is below $3.0\,\sigma$, the upper limits on the signal yields, including the additive and multiplicative systematic uncertainties, are determined at the 90\% C.L. using the Bayesian method~\cite{Zhu:2008ca}. 

For the additive systematic uncertainties, the likelihood scan is repeated with the nominal fit model modified according to each additive systematic variation. The likelihood distribution yielding the most conservative result is selected and subsequently used to account for multiplicative systematic uncertainties. 
For multiplicative systematic uncertainties, the selected likelihood distribution is additionally convolved with a Gaussian function that models the detection-efficiency uncertainty, centered at the nominal detection efficiency and with standard deviation $\sigma_{\varepsilon}$. Here, $\sigma_{\varepsilon}$ is defined as the product of the detection efficiency and the total relative multiplicative systematic uncertainty discussed in Sec.~\ref{SYS_UN}. The resulting likelihood distribution is used to determine the 90\% C.L. upper limit on the signal yield from the condition $\int_{0}^{N^{\rm UL}} L\,{\rm d}$$N_{\rm obs}/\int_{0}^{\infty} L\,{\rm d}$$N_{\rm obs}=0.9$. The corresponding cross section upper limits $(\sigma^{\rm UL})$ are derived accordingly.

\subsection{Determination of Born cross section}
At a given energy point, the Born cross section for the reactions $\KXX$ is calculated by 
\begin{linenomath*}
\begin{equation}
\sigma^{B} =\frac{N_{\rm obs}}{{\cal{L}}\cdot(1 + \delta)\cdot\frac{1}{|1 - \prod|^{2}}\cdot\varepsilon\cdot{\cal B}_{\rm inter}},
\label{bcs}
\end{equation}
\end{linenomath*}
where $N_{\rm obs}$ denotes the number of observed signal events in data, $\mathcal{L}$ is the integrated luminosity, $(1+\delta)$ is the ISR correction factor, $\frac{1}{|1-\Pi|^{2}}$ is the vacuum polarization (VP) correction factor, $\varepsilon$ is the detection efficiency and ${\cal B}_{\rm inter}$ is the product of the branching fractions of the intermediate states ($\KS \ar\pi^+\pi^-,\ \bar\Xi^0\ar\pi^0\bar\Lambda,\ \bar\Lambda\ar \bar p\pi^+\ \text{and}\ \pi^0 \ar \gamma \gamma$) taken from the PDG~\cite{PDG2020}. The VP correction factor is calculated according to ref.~\cite{Jegerlehner:2011ti}. The ISR correction factor is obtained using the Quantum Electrodynamics calculation as described in ref.~\cite{Kuraev:1985hb}. 
The detection efficiencies and ISR correction factors are determined iteratively following the procedure proposed in ref.~\cite{Sun:2020ehv} to ensure an accurate measurement of the Born cross sections. 
\begin{table}[!htp]
  \centering
  \caption{\small
  Numerical results for $\KXL$, where $\frac{1}{|1 - \prod|^{2}}$ is the VP correction factor, (1+$\delta$)$\cdot \varepsilon$ is the product of the ISR correction factor and the detection efficiency, $N_{\rm obs}$ denotes the number of the signal events, $N^{\rm UL}$ is the upper limit on the signal yield, $\sigma^{B}$ represents the Born cross section, and $\sigma^{\rm UL}$ is the upper limit on the Born cross section, taking systematic uncertainties into account. The first and second uncertainties for $\sigma^{B}$ are statistical and systematic, respectively. The $\cal{S}~(\sigma)$ is the statistical significance.
  }
   \scalebox{0.69}{
  \begin{tabular}{cccclllc}  \hline \hline
  \multicolumn{1}{c}{$\sqrt{s}$ (GeV)} & \multicolumn{1}{c}{${\cal L}$ (pb$^{-1})$} & \multicolumn{1}{c}{$\frac{1}{|1 - \prod|^{2}}$}&\multicolumn{1}{c}{(1+$\delta$)$\cdot \varepsilon~(\%)$} & \multicolumn{1}{c}{$N_{\rm obs}$ ($N^{\rm UL}$)}&\multicolumn{1}{c}{$\sigma^{B}$ (fb)} &\multicolumn{1}{c}{$\sigma^{\rm UL}$ (fb)} & $\cal{S}~(\sigma)$\\ \hline
3.51000 	&404.7	&1.04	&8.4	&$17.6^{+9.3}_{-8.3}~(\textless 31.6)$	&$1129^{+598}_{-532} \pm 72$	&$\textless 2031$	&1.6	\\
3.51440 	&40.9	&1.04	&8.6	&$0.0^{+1.6}_{0.0}~(\textless 5.8)$	&$0^{+1022}_{-0} \pm 0$	&$\textless 3622$	&0.0	\\
3.55281 	&42.2	&1.04	&8.4	&$0.0^{+2.6}_{-1.7}~(\textless 5.8)$	&$5^{+1586}_{-1050} \pm 0$	&$\textless 3576$	&0.0	\\
3.55400 	&129.4	&1.04	&8.6	&$8.3^{+5.5}_{-4.6}~(\textless 16.3)$	&$1638^{+1094}_{-912} \pm 104$	&$\textless 3229$	&1.4	\\
3.58154 	&85.7	&1.04	&8.3	&$4.1^{+4.8}_{-2.8}~(\textless 9.6)$	&$1281^{+1481}_{-870} \pm 81$	&$\textless 2986$	&2.4	\\
3.65000 	&410	&1.02	&8.6	&$16.9^{+7.4}_{-6.5}~(\textless 26.5)$	&$1085^{+474}_{-419} \pm 69$	&$\textless 1699$	&2.1	\\
3.67020 	&83.6	&0.99	&8.5	&$7.9^{+3.7}_{-2.9}~(\textless 13.0)$	&$2589^{+1191}_{-933} \pm 164$	&$\textless 4241$	&1.7	\\
3.76800 	&415.8	&1.05	&9.1	&$0.0^{+2.8}_{0.0}~(\textless 8.1)$	&$0^{+160}_{-0} \pm 0$	&$\textless 464$	&0.0	\\
3.77300 	&20274.8	&1.06	&9.3	&$311.4^{+47.5}_{-46.6}~(\textless 50.0)$	&$356^{+54}_{-53} \pm 23$	&$\textless 57$	&5.0	\\
3.78000 	&410	&1.06	&9.2	&$4.9^{+7.1}_{-6.1}~(\textless 15.5)$	&$282^{+406}_{-350} \pm 18$	&$\textless 886$	&0.6	\\
3.80765 	&50.5	&1.06	&9.7	&$4.1^{+2.9}_{-2.0}~(\textless 8.3)$	&$1801^{+1263}_{-880} \pm 114$	&$\textless 3656$	&1.1	\\
3.86741 	&108.9	&1.05	&9.8	&$2.9^{+3.7}_{-3.0}~(\textless 8.6)$	&$585^{+763}_{-603} \pm 37$	&$\textless 1753$	&2.2	\\
3.87131 	&110.3	&1.05	&9.8	&$1.6^{+3.4}_{-2.7}~(\textless 7.3)$	&$332^{+686}_{-541} \pm 21$	&$\textless 1476$	&0.4	\\
3.89600 	&52.6	&1.05	&9.8	&$3.2^{+3.4}_{-2.7}~(\textless 8.5)$	&$1355^{+1423}_{-1152} \pm 86$	&$\textless 3594$	&1.1	\\
4.00762 	&482	&1.04	&10.2	&$3.6^{+7.5}_{-6.7}~(\textless 19.2)$	&$161^{+332}_{-299} \pm 10$	&$\textless 855$	&0.4	\\
4.08545 	&52.9	&1.05	&10.1	&$2.2^{+2.1}_{-1.3}~(\textless 5.3)$	&$914^{+851}_{-532} \pm 58$	&$\textless 2170$	&1.5	\\
4.12848 	&401.5	&1.05	&9.4	&$1.3^{+4.6}_{-3.8}~(\textless 9.2)$	&$77^{+267}_{-221} \pm 5$	&$\textless 531$	&0.2	\\
4.15744 	&408.7	&1.05	&9.6	&$3.9^{+5.8}_{-4.7}~(\textless 13.8)$	&$217^{+320}_{-262} \pm 14$	&$\textless 768$	&0.6	\\
4.17800 	&3194.5	&1.05	&9.9	&$6.3^{+13.8}_{-12.9}~(\textless 26.8)$	&$44^{+95}_{-89} \pm 3$	&$\textless 184$	&0.3	\\
4.18880 	&526.7	&1.06	&10.1	&$6.4^{+14.9}_{-7.3}~(\textless 16.6)$	&$263^{+611}_{-300} \pm 17$	&$\textless 679$	&0.8	\\
4.19890 	&526	&1.06	&10.3	&$0.0^{+2.0}_{0.0}~(\textless 6.9)$	&$0^{+78}_{-0} \pm 0$	&$\textless 277$	&0.0	\\
4.20920 	&572.1	&1.06	&10.2	&$0.0^{+5.1}_{0.0}~(\textless 10.0)$	&$0^{+190}_{-0} \pm 0$	&$\textless 372$	&0.0	\\
4.21870 	&569.2	&1.06	&10.2	&$8.6^{+6.9}_{-6.1}~(\textless 19.4)$	&$323^{+258}_{-227} \pm 20$	&$\textless 724$	&1.0	\\
4.22626 	&1100.9	&1.06	&10.3	&$10.0^{+8.8}_{-8.0}~(\textless 23.3)$	&$193^{+168}_{-153} \pm 12$	&$\textless 447$	&0.9	\\
4.23570 	&530.3	&1.06	&10.1	&$9.9^{+6.3}_{-5.5}~(\textless 18.8)$	&$401^{+255}_{-223} \pm 25$	&$\textless 763$	&1.3	\\
4.24166 	&55.9	&1.06	&10.3	&$0.5^{+1.7}_{-1.1}~(\textless 3.9)$	&$203^{+636}_{-409} \pm 13$	&$\textless 1477$	&0.3	\\
4.24380 	&538.1	&1.06	&10.2	&$1.1^{+5.9}_{-5.1}~(\textless 11.1)$	&$44^{+231}_{-200} \pm 3$	&$\textless 438$	&0.1	\\
4.25797 	&828.4	&1.05	&10.3	&$11.8^{+8.2}_{-7.5}~(\textless 23.7)$	&$300^{+209}_{-191} \pm 19$	&$\textless 603$	&1.1	\\
4.26680 	&531.1	&1.05	&10.1	&$0.0^{+5.6}_{0.0}~(\textless 10.0)$	&$0^{+227}_{-0} \pm 0$	&$\textless 407$	&0.0	\\
4.27770 	&175.7	&1.05	&10.1	&$2.3^{+3.5}_{-2.7}~(\textless 7.8)$	&$288^{+430}_{-336} \pm 18$	&$\textless 960$	&0.6	\\
4.28788 	&502.4	&1.05	&9.4	&$0.0^{+4.1}_{0.0}~(\textless 8.0)$	&$0^{+188}_{-0} \pm 0$	&$\textless 369$	&0.0	\\
4.30789 	&45.1	&1.05	&10.5	&$1.0^{+1.5}_{-1.5}~(\textless 3.8)$	&$445^{+706}_{-692} \pm 28$	&$\textless 1754$	&0.7	\\
4.31205 	&501.2	&1.05	&9.5	&$2.7^{+6.2}_{-5.3}~(\textless 12.7)$	&$126^{+284}_{-244} \pm 8$	&$\textless 584$	&0.4	\\
4.33739 	&505	&1.05	&9.4	&$0.0^{+3.1}_{0.0}~(\textless 7.7)$	&$0^{+143}_{-0} \pm 0$	&$\textless 352$	&0.0	\\
4.35826 	&543.9	&1.05	&10.3	&$9.0^{+6.3}_{-5.6}~(\textless 17.9)$	&$347^{+245}_{-215} \pm 22$	&$\textless 694$	&1.2	\\
4.37737 	&522.7	&1.05	&9.6	&$2.5^{+5.5}_{-4.7}~(\textless 11.5)$	&$110^{+240}_{-204} \pm 7$	&$\textless 499$	&0.4	\\
4.38740 	&55.6	&1.05	&10.4	&$0.0^{+0.6}_{0.0}~(\textless 2.6)$	&$0^{+226}_{-0} \pm 0$	&$\textless 980$	&0.0	\\
4.39645 	&507.8	&1.05	&9.5	&$2.2^{+5.4}_{-4.7}~(\textless 10.8)$	&$101^{+245}_{-212} \pm 6$	&$\textless 488$	&0.3	\\
4.41558 	&1090.7	&1.05	&10.3	&$27.8^{+8.6}_{-7.9}~(\textless 38.7)$	&$541^{+168}_{-153} \pm 34$	&$\textless 754$	&2.8	\\
4.43624 	&569.9	&1.05	&9.5	&$1.9^{+5.1}_{-4.3}~(\textless 10.7)$	&$77^{+205}_{-173} \pm 5$	&$\textless 430$	&0.3	\\
4.46706 	&111.1	&1.05	&10.3	&$1.9^{+3.1}_{-2.1}~(\textless 7.0)$	&$362^{+580}_{-406} \pm 23$	&$\textless 1325$	&0.6	\\
4.52714 	&112.1	&1.05	&10.2	&$0.0^{+0.8}_{0.0}~(\textless 4.0)$	&$0^{+153}_{-0} \pm 0$	&$\textless 761$	&0.0	\\
4.57450 	&48.9	&1.05	&10.2	&$1.3^{+1.7}_{-0.9}~(\textless 4.2)$	&$570^{+753}_{-411} \pm 36$	&$\textless 1837$	&1.0	\\
4.59953 	&586.9	&1.05	&10.3	&$1.2^{+5.7}_{-4.6}~(\textless 9.2)$	&$43^{+206}_{-166} \pm 3$	&$\textless 332$	&0.1	\\
4.61186 	&103.7	&1.05	&9.2	&$1.5^{+1.8}_{-1.0}~(\textless 4.4)$	&$333^{+415}_{-232} \pm 21$	&$\textless 1004$	&1.0	\\
4.62800 	&521.5	&1.05	&9.1	&$10.2^{+6.6}_{-5.7}~(\textless 19.6)$	&$466^{+303}_{-258} \pm 29$	&$\textless 893$	&1.4	\\
4.64091 	&551.7	&1.05	&9.4	&$3.9^{+4.7}_{-3.9}~(\textless 10.7)$	&$163^{+199}_{-165} \pm 10$	&$\textless 451$	&0.7	\\
4.66124 	&529.4	&1.05	&9.1	&$4.3^{+6.9}_{-6.2}~(\textless 15.1)$	&$195^{+309}_{-277} \pm 12$	&$\textless 681$	&1.9	\\
4.68192 	&1667.4	&1.05	&9.2	&$0.0^{+6.0}_{0.0}~(\textless 14.0)$	&$0^{+84}_{-0} \pm 0$	&$\textless 198$	&0.0	\\
4.69882 	&535.5	&1.05	&9.2	&$8.5^{+6.5}_{-5.9}~(\textless 17.6)$	&$378^{+289}_{-263} \pm 24$	&$\textless 781$	&1.0	\\
4.73970 	&163.9	&1.05	&9.8	&$0.0^{+0.7}_{0.0}~(\textless 1.6)$	&$0^{+89}_{-0} \pm 0$	&$\textless 217$	&0.0	\\
4.75005 	&366.6	&1.05	&10.1	&$2.2^{+2.5}_{-2.3}~(\textless 5.9)$	&$130^{+145}_{-137} \pm 8$	&$\textless 348$	&0.5	\\
4.78054 	&511.5	&1.06	&10.1	&$9.6^{+5.9}_{-5.4}~(\textless 17.0)$	&$405^{+249}_{-226} \pm 25$	&$\textless 717$	&1.3	\\
4.84307 	&525.2	&1.06	&9.9	&$1.1^{+2.4}_{-1.5}~(\textless 5.5)$	&$47^{+100}_{-62} \pm 3$	&$\textless 230$	&0.3	\\
4.91802 	&207.8	&1.06	&9.7	&$0.0^{+0.7}_{0.0}~(\textless 3.1)$	&$0^{+77}_{-0} \pm 0$	&$\textless 333$	&0.0	\\
4.95093 	&159.3	&1.06	&9.7	&$1.2^{+2.1}_{-2.3}~(\textless 5.3)$	&$171^{+302}_{-324} \pm 11$	&$\textless 747$	&0.3	\\
  \hline\hline
  \end{tabular}
  }
  \label{tab:NUM_BCSL}
  \end{table}

  \begin{table}[!htp]
  \centering
  \caption{\small
  Numerical results for $\KXS$, where $\frac{1}{|1 - \prod|^{2}}$ is the VP correction factor, (1+$\delta$)$\cdot \varepsilon$ is the product of the ISR correction factor and the detection efficiency, $N_{\rm obs}$ denotes the number of the signal events, $N^{\rm UL}$ is the upper limit on the signal yield, $\sigma^{B}$ represents the Born cross section, and $\sigma^{\rm UL}$ is the upper limit on the Born cross section taking the systematic uncertainties into account. The first and second uncertainties for $\sigma^{B}$ are statistical and systematic, respectively. The $\cal{S}~(\sigma)$ is the statistical significance.
  }
   \scalebox{0.69}{
  \begin{tabular}{cccclllc}  \hline \hline
    \multicolumn{1}{c}{$\sqrt{s}$ (GeV)} & \multicolumn{1}{c}{${\cal L}$ (pb$^{-1})$} & \multicolumn{1}{c}{$\frac{1}{|1 - \prod|^{2}}$}&\multicolumn{1}{c}{(1+$\delta$)$\cdot \varepsilon~(\%)$} & \multicolumn{1}{c}{$N_{\rm obs}$ ($N^{\rm UL}$)}&\multicolumn{1}{c}{$\sigma^{B}$ (fb)} &\multicolumn{1}{c}{$\sigma^{\rm UL}$ (fb)} & $\cal{S}~(\sigma)$\\ \hline
3.51000 	&404.7	&1.04	&7.9	&$27.0^{+10.8}_{-9.8}~(\textless 41.3)$	&$1863^{+749}_{-679} \pm 102$	&$\textless 2852$	&2.2	\\
3.51440 	&40.9	&1.04	&8	&$0.2^{+2.7}_{0.0}~(\textless 5.7)$	&$140^{+1831}_{-0} \pm 8$	&$\textless 3836$	&0.1	\\
3.55281 	&42.2	&1.04	&7.8	&$2.1^{+4.4}_{-2.9}~(\textless 9.5)$	&$1377^{+2914}_{-1949} \pm 76$	&$\textless 6360$	&0.5	\\
3.55400 	&129.4	&1.04	&7.6	&$6.0^{+5.6}_{-4.4}~(\textless 14.0)$	&$1336^{+1238}_{-991} \pm 73$	&$\textless 3120$	&1.0	\\
3.58154 	&85.7	&1.04	&7.5	&$1.9^{+4.4}_{-2.6}~(\textless 7.7)$	&$642^{+1509}_{-872} \pm 35$	&$\textless 2628$	&0.5	\\
3.65000 	&410	&1.02	&7.6	&$25.8^{+9.5}_{-8.5}~(\textless 38.4)$	&$1855^{+681}_{-610} \pm 102$	&$\textless 2763$	&2.5	\\
3.67020 	&83.6	&0.99	&7.8	&$0.0^{+1.0}_{0.0}~(\textless 4.1)$	&$0^{+342}_{-0} \pm 0$	&$\textless 1455$	&0.0	\\
3.76800 	&415.8	&1.05	&8.5	&$10.6^{+8.2}_{-7.1}~(\textless 22.0)$	&$650^{+503}_{-438} \pm 36$	&$\textless 1347$	&1.1	\\
3.77300 	&20274.8	&1.06	&8.7	&$572.3^{+55.9}_{-55.2}~(\textless 50.0)$	&$702^{+69}_{-68} \pm 39$	&$\textless 61$	&8.1	\\
3.78000 	&410	&1.06	&8.6	&$9.6^{+8.8}_{-7.7}~(\textless 22.0)$	&$584^{+538}_{-472} \pm 32$	&$\textless 1343$	&0.9	\\
3.80765 	&50.5	&1.06	&8.9	&$0.3^{+3.2}_{-2.1}~(\textless 6.6)$	&$131^{+1540}_{-1007} \pm 7$	&$\textless 3195$	&0.1	\\
3.86741 	&108.9	&1.05	&9	&$1.8^{+3.7}_{-2.8}~(\textless 8.0)$	&$408^{+833}_{-627} \pm 22$	&$\textless 1783$	&0.4	\\
3.87131 	&110.3	&1.05	&8.9	&$3.4^{+4.7}_{-3.8}~(\textless 12.4)$	&$753^{+1043}_{-854} \pm 41$	&$\textless 2750$	&0.6	\\
3.89600 	&52.6	&1.05	&9.1	&$2.9^{+2.5}_{-1.7}~(\textless 6.5)$	&$1334^{+1126}_{-795} \pm 73$	&$\textless 2970$	&1.3	\\
4.00762 	&482	&1.04	&9.6	&$3.4^{+7.1}_{-6.5}~(\textless 16.0)$	&$161^{+338}_{-310} \pm 9$	&$\textless 760$	&0.4	\\
4.08545 	&52.9	&1.05	&9.4	&$1.5^{+2.3}_{-1.6}~(\textless 5.3)$	&$639^{+1007}_{-703} \pm 35$	&$\textless 2321$	&0.6	\\
4.12848 	&401.5	&1.05	&8.8	&$18.7^{+8.2}_{-7.2}~(\textless 29.8)$	&$1147^{+502}_{-443} \pm 62$	&$\textless 1827$	&2.0	\\
4.15744 	&408.7	&1.05	&8.9	&$8.2^{+7.1}_{-6.1}~(\textless 20.5)$	&$492^{+425}_{-366} \pm 27$	&$\textless 1233$	&1.0	\\
4.17800 	&3194.5	&1.05	&9.4	&$36.3^{+18.0}_{-17.1}~(\textless 59.6)$	&$264^{+131}_{-124} \pm 14$	&$\textless 433$	&1.6	\\
4.18880 	&526.7	&1.06	&9.4	&$0.0^{+3.7}_{0.0}~(\textless 9.9)$	&$0^{+160}_{-0} \pm 0$	&$\textless 434$	&0.0	\\
4.19890 	&526	&1.06	&9.5	&$15.0^{+8.9}_{-7.9}~(\textless 26.9)$	&$649^{+384}_{-340} \pm 35$	&$\textless 1165$	&1.4	\\
4.20920 	&572.1	&1.06	&9.5	&$11.2^{+7.5}_{-6.6}~(\textless 21.3)$	&$444^{+297}_{-262} \pm 24$	&$\textless 848$	&1.3	\\
4.21870 	&569.2	&1.06	&9.5	&$23.5^{+9.5}_{-8.1}~(\textless 34.7)$	&$947^{+382}_{-327} \pm 52$	&$\textless 1399$	&2.3	\\
4.22626 	&1100.9	&1.06	&9.7	&$19.9^{+9.7}_{-8.8}~(\textless 33.9)$	&$403^{+196}_{-178} \pm 22$	&$\textless 686$	&1.7	\\
4.23570 	&530.3	&1.06	&9.4	&$11.8^{+6.2}_{-5.4}~(\textless 20.0)$	&$517^{+273}_{-237} \pm 28$	&$\textless 874$	&1.7	\\
4.24166 	&55.9	&1.06	&9.8	&$3.5^{+2.6}_{-1.8}~(\textless 7.3)$	&$1375^{+1017}_{-696} \pm 75$	&$\textless 2898$	&1.1	\\
4.24380 	&538.1	&1.06	&9.4	&$20.6^{+8.1}_{-7.2}~(\textless 31.2)$	&$885^{+347}_{-309} \pm 48$	&$\textless 1343$	&2.3	\\
4.25797 	&828.4	&1.05	&9.7	&$12.4^{+8.2}_{-7.4}~(\textless 23.6)$	&$337^{+222}_{-200} \pm 18$	&$\textless 639$	&1.2	\\
4.26680 	&531.1	&1.05	&9.5	&$14.9^{+8.7}_{-6.6}~(\textless 24.0)$	&$639^{+373}_{-285} \pm 35$	&$\textless 1033$	&1.7	\\
4.27770 	&175.7	&1.05	&9.4	&$2.7^{+3.2}_{-2.5}~(\textless 7.7)$	&$360^{+425}_{-331} \pm 20$	&$\textless 1018$	&0.8	\\
4.28788 	&502.4	&1.05	&8.9	&$12.0^{+6.8}_{-6.1}~(\textless 21.2)$	&$582^{+332}_{-296} \pm 32$	&$\textless 1030$	&1.5	\\
4.30789 	&45.1	&1.05	&9.9	&$2.0^{+1.9}_{-1.2}~(\textless 4.9)$	&$994^{+932}_{-586} \pm 54$	&$\textless 2395$	&0.6	\\
4.31205 	&501.2	&1.05	&8.8	&$9.2^{+7.9}_{-7.0}~(\textless 22.0)$	&$457^{+391}_{-346} \pm 25$	&$\textless 1090$	&1.0	\\
4.33739 	&505	&1.05	&8.8	&$2.9^{+6.4}_{-5.4}~(\textless 13.2)$	&$144^{+311}_{-264} \pm 8$	&$\textless 646$	&0.4	\\
4.35826 	&543.9	&1.05	&9.7	&$13.1^{+7.5}_{-6.6}~(\textless 23.2)$	&$542^{+309}_{-272} \pm 30$	&$\textless 957$	&1.5	\\
4.37737 	&522.7	&1.05	&8.9	&$4.9^{+6.2}_{-5.5}~(\textless 14.3)$	&$231^{+292}_{-256} \pm 13$	&$\textless 669$	&0.6	\\
4.38740 	&55.6	&1.05	&9.7	&$2.0^{+1.8}_{-1.1}~(\textless 4.6)$	&$810^{+717}_{-446} \pm 44$	&$\textless 1865$	&2.8	\\
4.39645 	&507.8	&1.05	&8.8	&$8.2^{+6.3}_{-5.6}~(\textless 16.8)$	&$402^{+308}_{-274} \pm 22$	&$\textless 821$	&1.1	\\
4.41558 	&1090.7	&1.05	&9.6	&$5.0^{+8.0}_{-7.2}~(\textless 17.3)$	&$104^{+168}_{-150} \pm 6$	&$\textless 362$	&0.5	\\
4.43624 	&569.9	&1.05	&8.9	&$8.1^{+6.1}_{-5.4}~(\textless 18.1)$	&$347^{+260}_{-230} \pm 19$	&$\textless 777$	&1.1	\\
4.46706 	&111.1	&1.05	&9.5	&$0.0^{+1.0}_{0.0}~(\textless 4.5)$	&$0^{+200}_{-0} \pm 0$	&$\textless 922$	&0.0	\\
4.52714 	&112.1	&1.05	&9.5	&$3.0^{+2.1}_{-2.0}~(\textless 6.1)$	&$611^{+423}_{-406} \pm 33$	&$\textless 1241$	&2.6	\\
4.57450 	&48.9	&1.05	&9.5	&$0.0^{+2.1}_{0.0}~(\textless 4.7)$	&$0^{+996}_{-0} \pm 0$	&$\textless 2187$	&0.0	\\
4.59953 	&586.9	&1.05	&9.7	&$8.4^{+7.1}_{-6.4}~(\textless 18.3)$	&$320^{+272}_{-243} \pm 17$	&$\textless 699$	&1.0	\\
4.61186 	&103.7	&1.05	&8.8	&$0.0^{+1.3}_{0.0}~(\textless 4.1)$	&$0^{+310}_{-0} \pm 0$	&$\textless 982$	&0.0	\\
4.62800 	&521.5	&1.05	&8.7	&$8.0^{+7.2}_{-6.4}~(\textless 18.2)$	&$382^{+345}_{-309} \pm 21$	&$\textless 871$	&0.9	\\
4.64091 	&551.7	&1.05	&8.7	&$0.0^{+2.1}_{0.0}~(\textless 6.6)$	&$0^{+95}_{-0} \pm 0$	&$\textless 298$	&0.0	\\
4.66124 	&529.4	&1.05	&8.6	&$4.4^{+6.6}_{-5.7}~(\textless 14.6)$	&$212^{+317}_{-273} \pm 12$	&$\textless 698$	&0.5	\\
4.68192 	&1667.4	&1.05	&8.7	&$0.0^{+5.3}_{0.0}~(\textless 13.1)$	&$0^{+80}_{-0} \pm 0$	&$\textless 197$	&0.0	\\
4.69882 	&535.5	&1.05	&8.8	&$5.6^{+6.8}_{-5.9}~(\textless 15.6)$	&$261^{+313}_{-273} \pm 14$	&$\textless 722$	&0.7	\\
4.73970 	&163.9	&1.05	&9.3	&$4.0^{+2.3}_{-1.7}~(\textless 7.3)$	&$571^{+334}_{-240} \pm 31$	&$\textless 1041$	&0.8	\\
4.75005 	&366.6	&1.05	&9.2	&$6.8^{+3.3}_{-4.2}~(\textless 11.4)$	&$437^{+211}_{-270} \pm 24$	&$\textless 734$	&0.4	\\
4.78054 	&511.5	&1.06	&9.3	&$3.5^{+4.2}_{-3.4}~(\textless 9.7)$	&$159^{+192}_{-155} \pm 9$	&$\textless 444$	&0.7	\\
4.84307 	&525.2	&1.06	&9.2	&$7.3^{+6.1}_{-5.2}~(\textless 15.5)$	&$326^{+272}_{-233} \pm 18$	&$\textless 696$	&1.0	\\
4.91802 	&207.8	&1.06	&9.1	&$2.7^{+4.2}_{-2.9}~(\textless 8.4)$	&$308^{+475}_{-332} \pm 17$	&$\textless 961$	&0.6	\\
4.95093 	&159.3	&1.06	&9.1	&$3.8^{+2.6}_{-3.6}~(\textless 7.7)$	&$569^{+396}_{-544} \pm 31$	&$\textless 1156$	&0.8	\\
  \hline\hline
  \end{tabular}
  }
  \label{tab:NUM_BCSS}
  \end{table}

\section{Systematic uncertainty}\label{SYS_UN}
The systematic uncertainties in measurements of Born cross sections arise from various sources, classified as multiplicative and additive. Multiplicative terms refer to uncertainties due to integrated luminosity, $K_S^0$ reconstruction, $\bar{\Xi}^0$ reconstruction, branching fraction, and input line shape. The additive terms include the signal and background shapes in the fit method.

\subsection{Luminosity}
The integrated luminosity is measured using Bhabha events. The corresponding uncertainties are 1.0\% below 4.0~GeV~\cite{ene1}, 0.7\% in the range from 4.0 to 4.6~GeV~\cite{BESIII:2022dxl}, and 0.6\% above 4.6~GeV~\cite{ene3}, which are taken as the systematic uncertainties associated with the luminosity measurement.

\subsection{$\KS$ reconstruction}
The systematic uncertainty related to the $K_S^0$ reconstruction is evaluated using control samples of the processes $J/\psi \to K^{*}(892)^{\pm}K^{\mp}$ and $K^{*}(892)^{\pm} \to K_S^0 \pi^{\pm}$~\cite{BESIII:2021kwf}. The effects arising from tracking and PID, the $K_S^0$ decay-length requirement, and the $K_S^0$ mass window are considered, and the resulting systematic uncertainty is estimated to be 2.1\%.

\subsection{$\Xib$ reconstruction}
The systematic uncertainty associated with the $\bar{\Xi}^{0}$ reconstruction arises from the tracking and PID, the $\pi^0/\bar{\Lambda}/\bar{\Xi}^{0}$ reconstruction, the $\bar{\Lambda}$ decay-length requirement, and the mass-window requirements of the $\bar{\Lambda}$ and $\bar{\Xi}^{0}$. The combined uncertainty is estimated with a control sample of $\psi(3686)\ar\Xi^{0}\bar{\Xi}^{0}$ using the same method described in ref.~\cite{BESIII:2024xixi}. The efficiency difference between the data and the MC simulation is 4.5\%, which is assigned as the systematic uncertainty.

\subsection{Branching fraction} 
The uncertainties related to the branching fraction of the intermediate decays $\KS \ar\pi^+\pi^-,\ \bar\Xi^0\ar\pi^0\bar\Lambda,\ \bar\Lambda\ar \bar p\pi^+\ \text{and}\ \pi^0 \ar \gamma \gamma$ are 0.8\%, using the PDG values~\cite{PDG2020}.

\subsection{Fit method}
Due to the limited statistics, the fit-related systematic uncertainties are evaluated by combining all data samples, including those associated with the signal and background shapes in the fit to the $M^{\rm Recoil}_{K_S^0\Xib}$ spectrum. The signal shape is changed from the signal MC shape to the MC shape convolved with a Gaussian function. The uncertainties due to the signal shapes are evaluated by replacing the nominal MC shapes of both signal processes with their respective Gaussian-convolved MC shapes. The resulting systematic uncertainties from the signal shape are 2.1\% and 1.4\% for $\KXL$ and $\KXS$, respectively. The background shape is adjusted from a second to a third-order polynomial. The corresponding systematic uncertainties are 3.0\% for $\KXL$ and 1.4\% for $\KXS$.

\subsection{Input line shape}
The systematic uncertainty for the input line shape of the cross section for determining the product
of the ISR correction and the detection efficiency $(1 + \delta)\cdot\varepsilon$ is estimated by varying the central value
of the nominal input line shape within $\pm1\sigma$ of the statistical uncertainty. The $(1 + \delta)\cdot\varepsilon$ values for each energy point are then recalculated. This process is repeated 100 times, and a Gaussian function is used to fit the $(1 + \delta)\cdot\varepsilon$ distribution. The width of the Gaussian function is taken as the corresponding systematic uncertainty.

\subsection{Total systematic uncertainty}
The various systematic uncertainties on the Born cross section measurements for the processes $\KXX$ are summarized in tables~\ref{systematic1} and~\ref{systematic2}. Assuming all sources are independent, the total systematic uncertainty is determined by adding these values in quadrature.
 \begin{table}[!hpt]
 	\begin{center}
 	{\caption{\small Systematic uncertainties (in \%) and their sources on the Born cross section measurement for the process $\KXL$. Here, Lum. denotes luminosity, $\KS$ Rec. and $\bar{\Xi}^{0}$ Rec. denote $\KS$ reconstruction and $\bar{\Xi}^{0}$ reconstruction, respectively, ${\cal{B}}$ denotes branching fraction, SS and BS denote signal and background shape, respectively, and ILS denotes input line shape.}
     \label{systematic1}
 	}
 	 \begin{tabular}{ccccccccc}\hline \hline
 $\sqrt{s}$ (GeV)            & Lum. &$K_S^0$ Rec. & $\bar{\Xi}^0$ Rec. & $\mathcal{B}$ & SS & BS & ILS &Total \\ \hline
 $\sqrt{s}<4.0$	            &1.0	&2.1	&4.5	&0.8	&2.1    &3.0	&1.0 	&6.3   \\
 $4.0\leq \sqrt{s}<4.6$	    &0.7	&2.1	&4.5	&0.8	&2.1	&3.0	&1.0 	&6.3	\\
 $\sqrt{s}>4.6$	            &0.6	&2.1	&4.5	&0.8	&2.1	&3.0	&1.0 	&6.3	 \\ 
 
 \hline \hline
   \end{tabular}
 	\end{center}
 \end{table}
\begin{table}[!hpt]
	\begin{center}
	{\caption{\small Systematic uncertainties (in \%) and their sources on the Born cross section measurement for the process $\KXS$. Here, Lum. denotes luminosity, $\KS$ Rec. and $\bar{\Xi}^{0}$ Rec. denote $\KS$ reconstruction and $\bar{\Xi}^{0}$ reconstruction, respectively, ${\cal{B}}$ denotes branching fraction, SS and BS denote signal and background shape, respectively, and ILS denotes input line shape.}
    \label{systematic2}
	}
	  \begin{tabular}{ccccccccc}\hline \hline
$\sqrt{s}$ (GeV)            & Lum. &$K_S^0$ Rec. & $\bar{\Xi}^0$ Rec. & $\mathcal{B}$ & SS & BS & ILS &Total \\ \hline 
$\sqrt{s}<4.0$	            &1.0	&2.1	&4.5	&0.8	&1.4	&1.4	&0.2 	&5.5	 	\\
$4.0\leq \sqrt{s}<4.6$	    &0.7	&2.1	&4.5	&0.8	&1.4	&1.4	&0.2 	&5.4		\\
$\sqrt{s}>4.6$	            &0.6    &2.1	&4.5	&0.8	&1.4	&1.4	&0.2 	&5.4		 	\\

\hline \hline
  \end{tabular}
	\end{center}
\end{table}
\newpage
 \section{Fit to the dressed cross section}
The potential resonances in the line shape of the cross sections for $\KXX$ are studied by fitting the dressed cross sections, $\sigma^{\rm dressed} =\sigma^{B}/|1-\Pi|^2$ with the least $\chi^{2}$ method defined as
\begin{linenomath*}
 \begin{equation}
\chi^{2} = \Delta X^{T}V^{-1}\Delta X,
 \end{equation}
\end{linenomath*}
where $V$ is the covariance matrix and $\Delta X$ is the vector of residuals between the measured and fitted cross sections. The covariance matrix $V$ incorporates the correlated and uncorrelated uncertainties among different energy points, where the systematic uncertainties due to the luminosity, $K_S^0$ reconstruction, $\bar{\Xi}^0$ reconstruction, and branching fraction are assumed to be fully correlated among the CM energies. The other sources of uncertainties are taken to be uncorrelated.

\begin{figure}[htbp]
  \centering
  \includegraphics[width=0.44\textwidth]{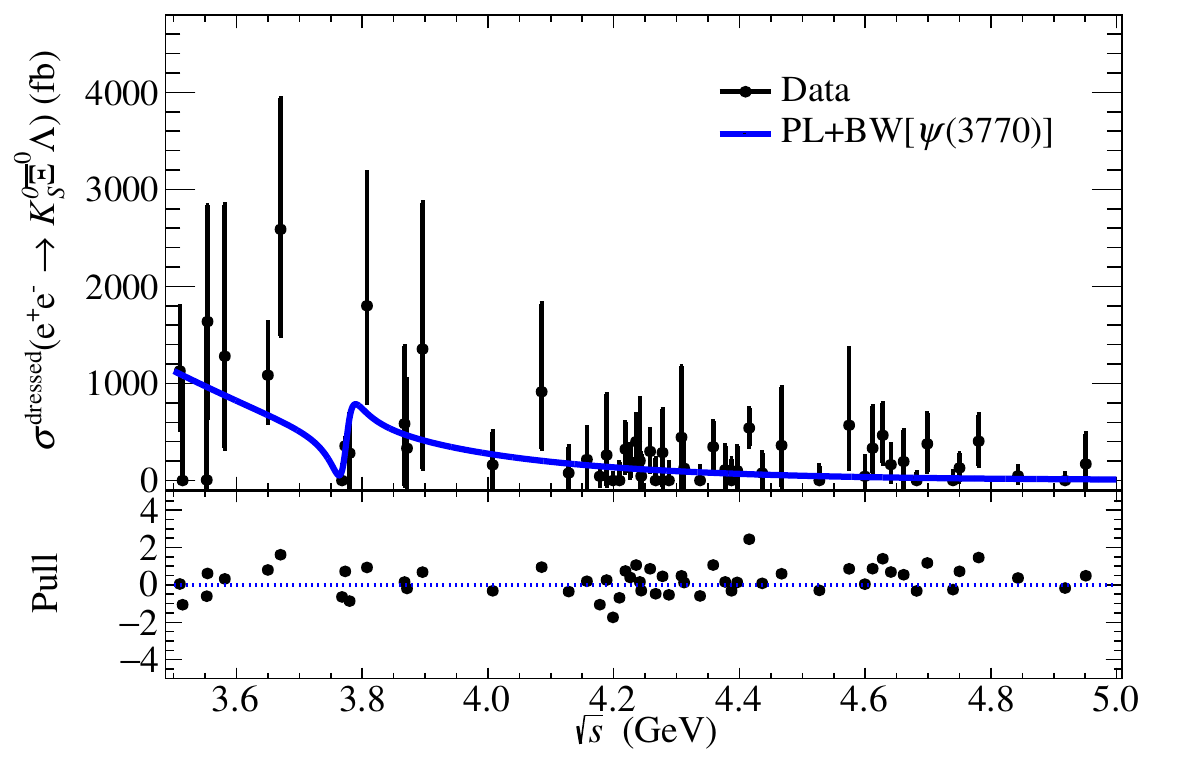}
  \includegraphics[width=0.44\textwidth]{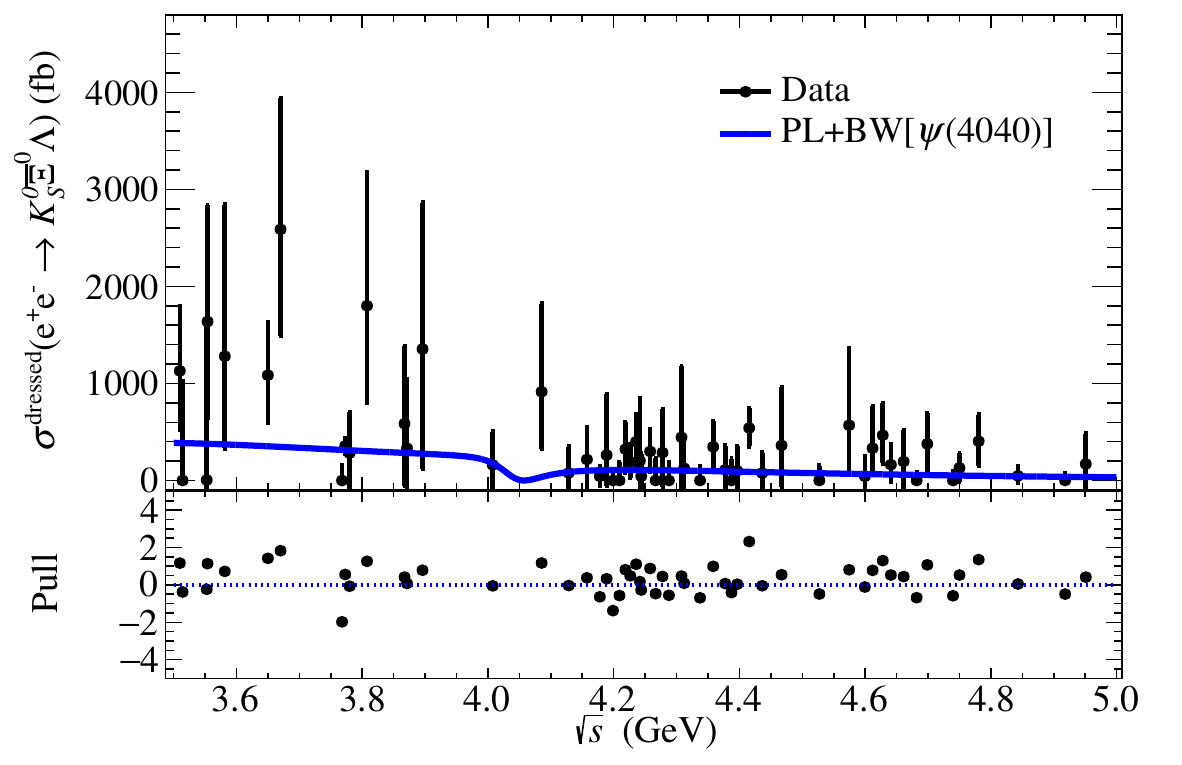}\\
  \includegraphics[width=0.44\textwidth]{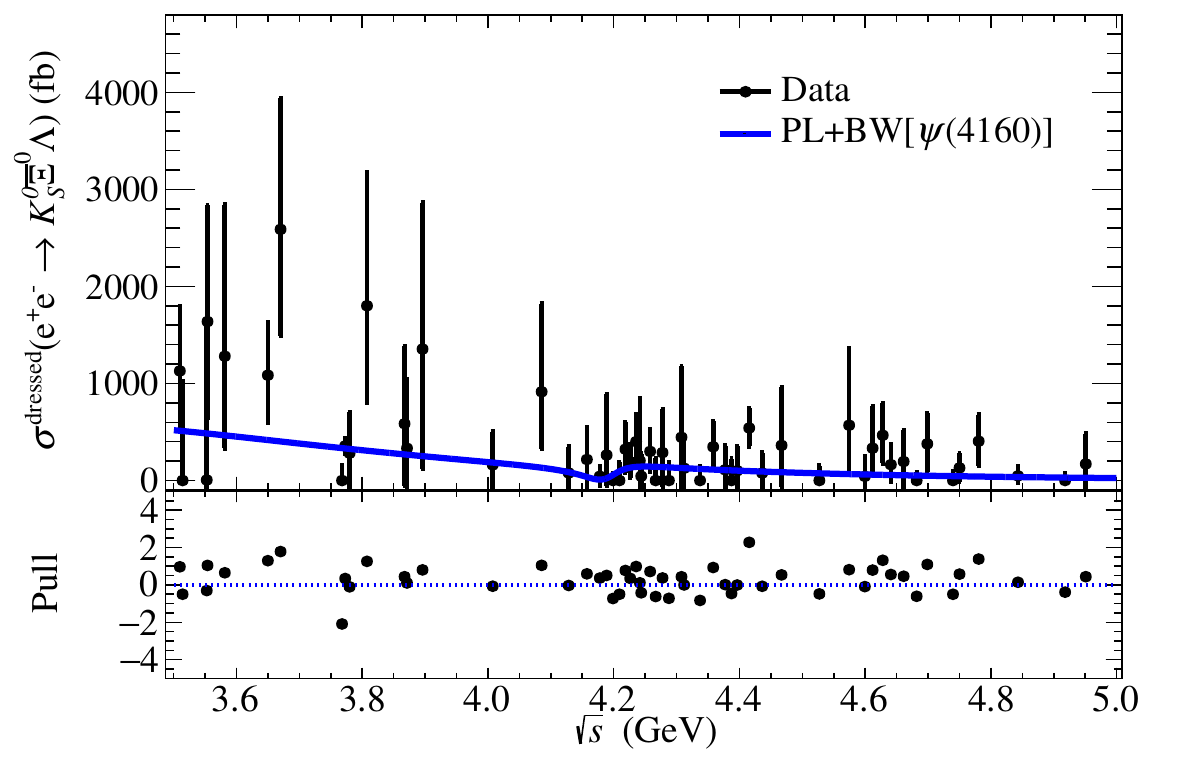}
  \includegraphics[width=0.44\textwidth]{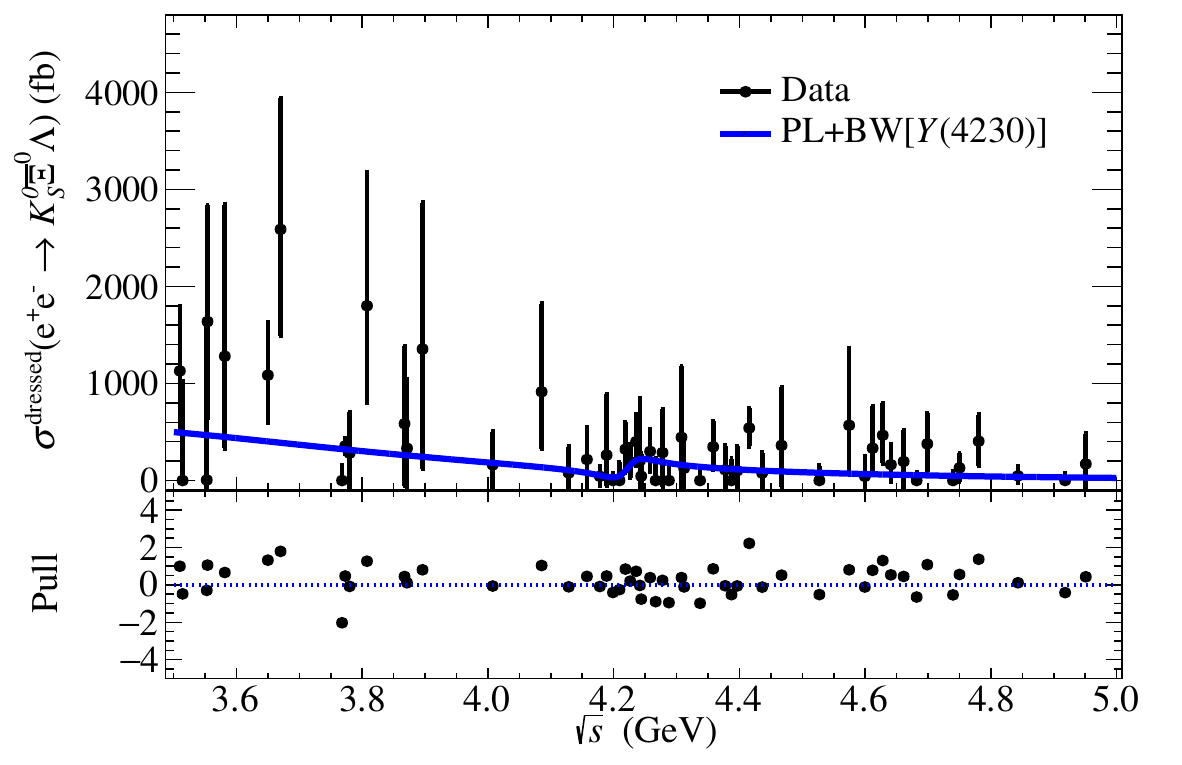}\\
  \includegraphics[width=0.44\textwidth]{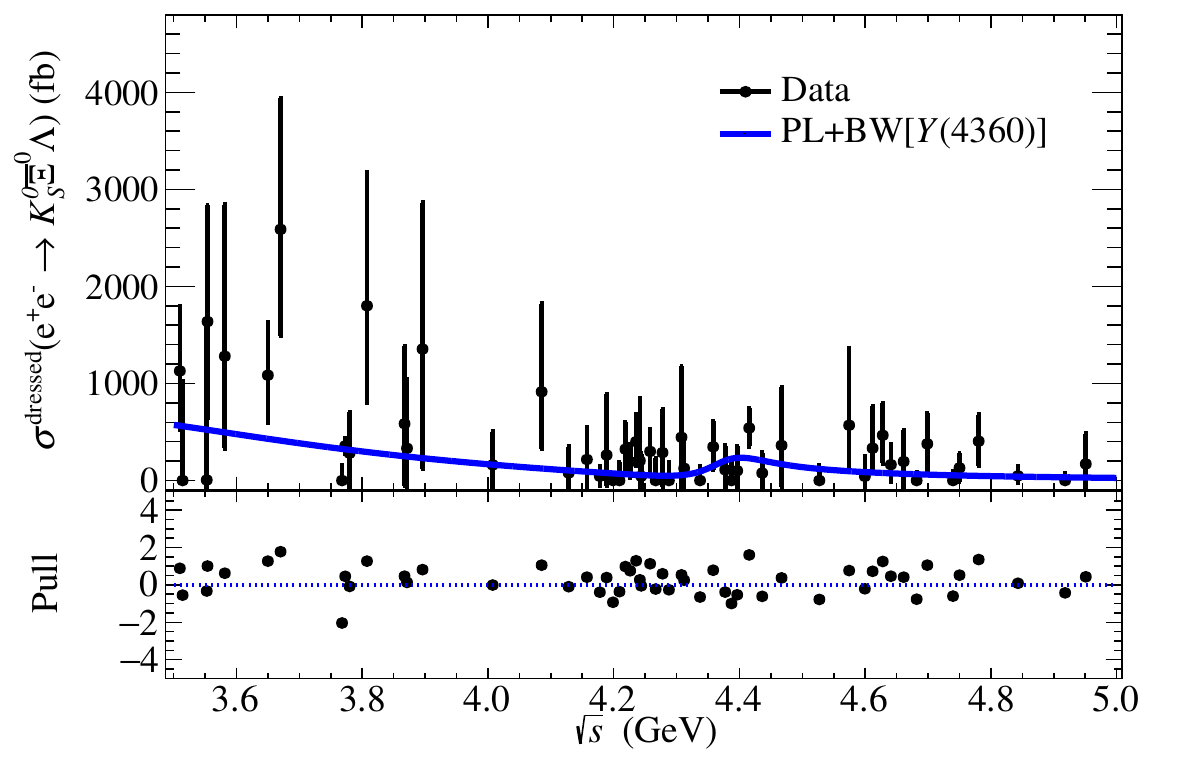}
  \includegraphics[width=0.44\textwidth]{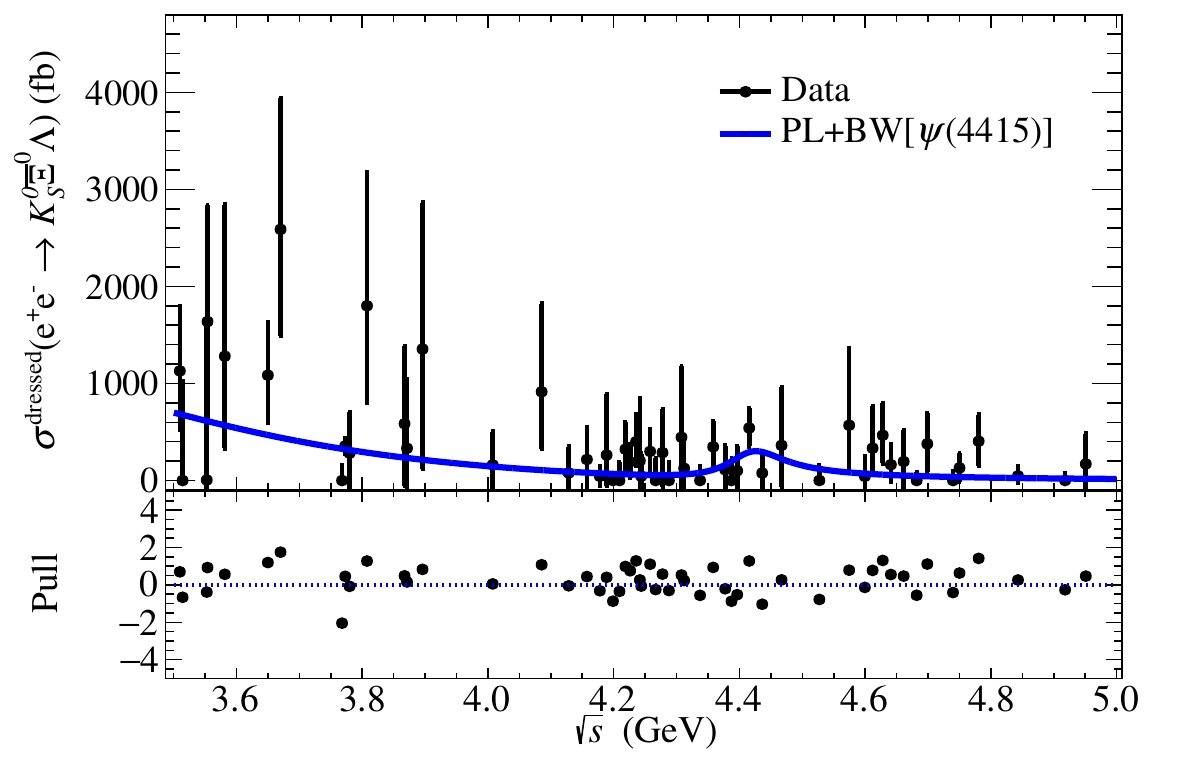}\\
  \includegraphics[width=0.44\textwidth]{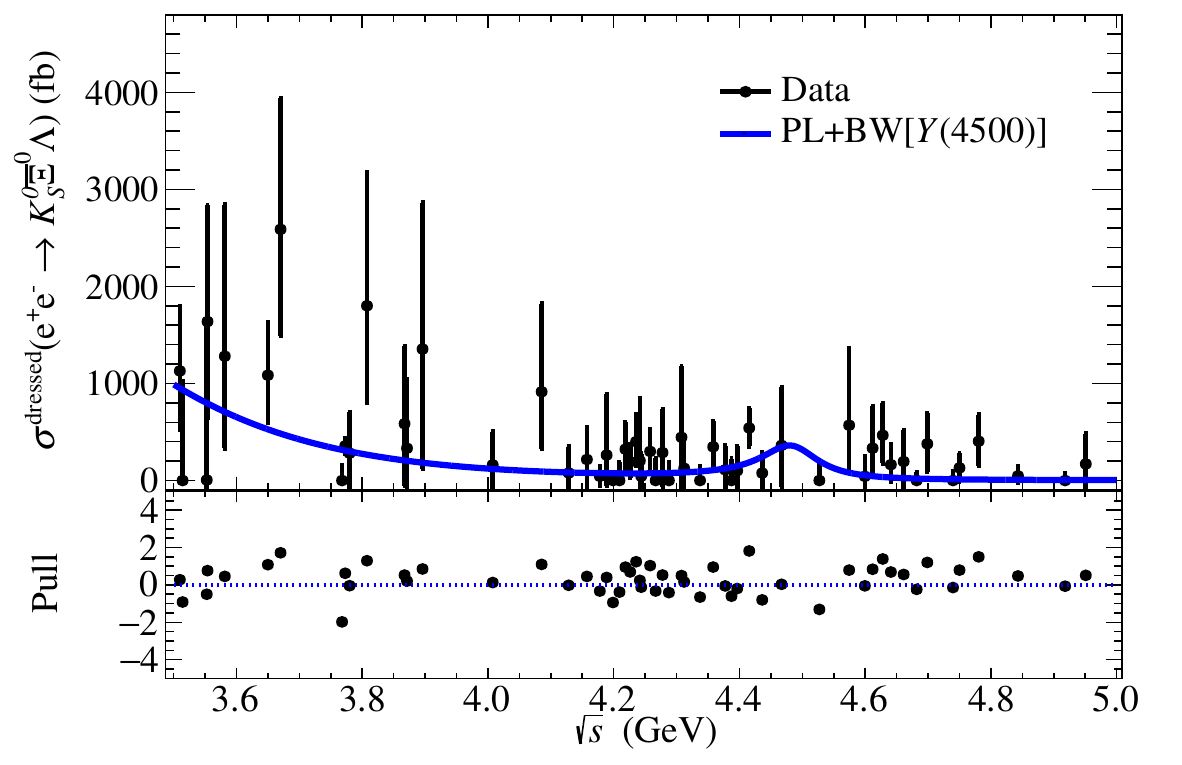}
  \includegraphics[width=0.44\textwidth]{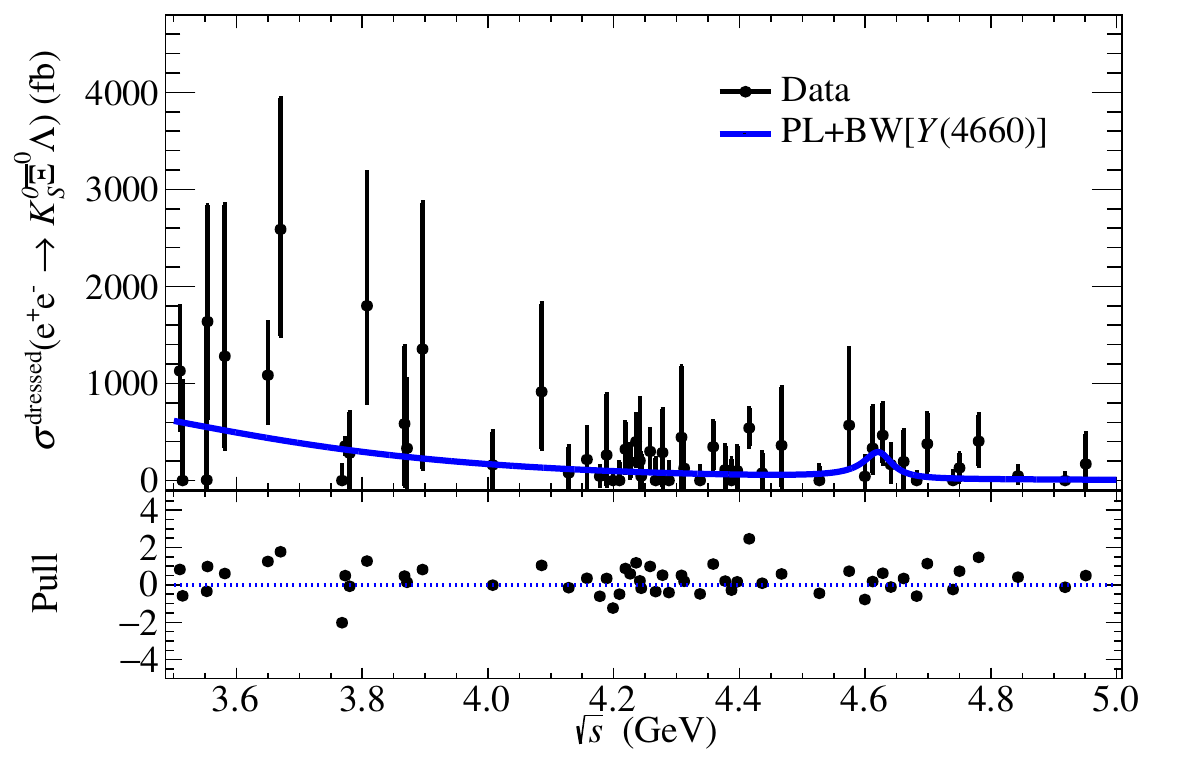}\\
  \includegraphics[width=0.44\textwidth]{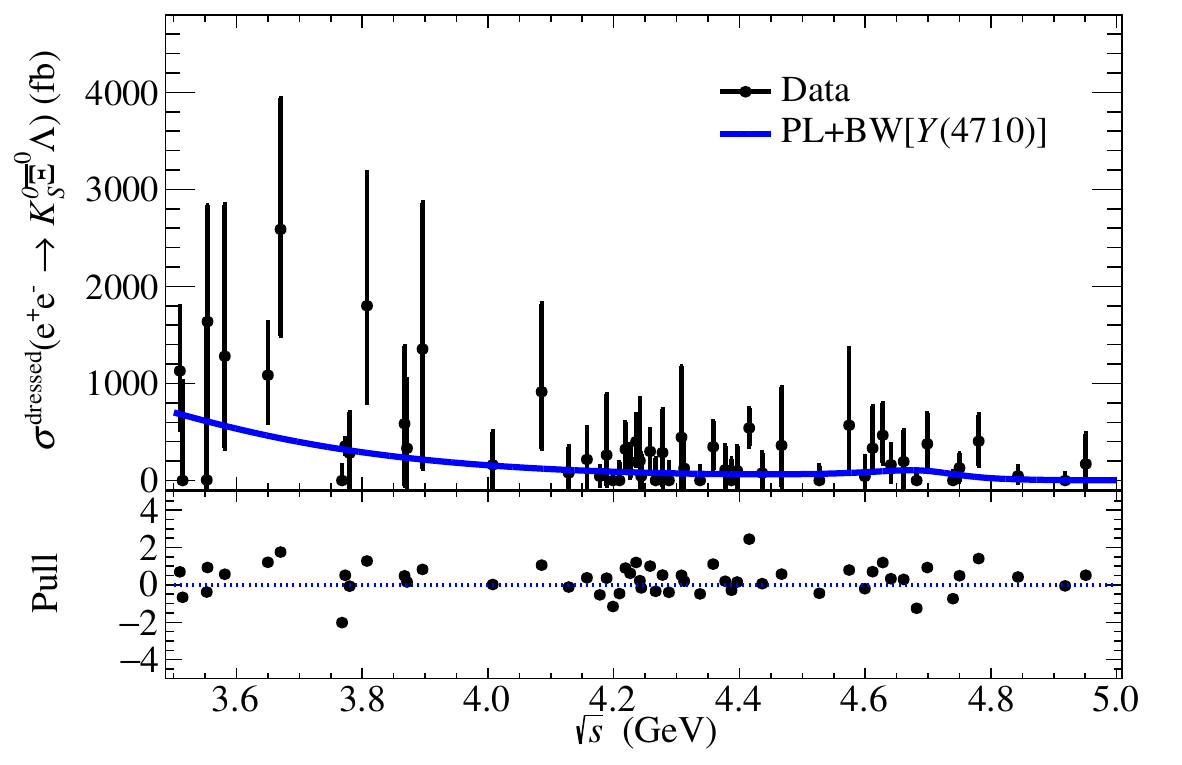}
  \includegraphics[width=0.44\textwidth]{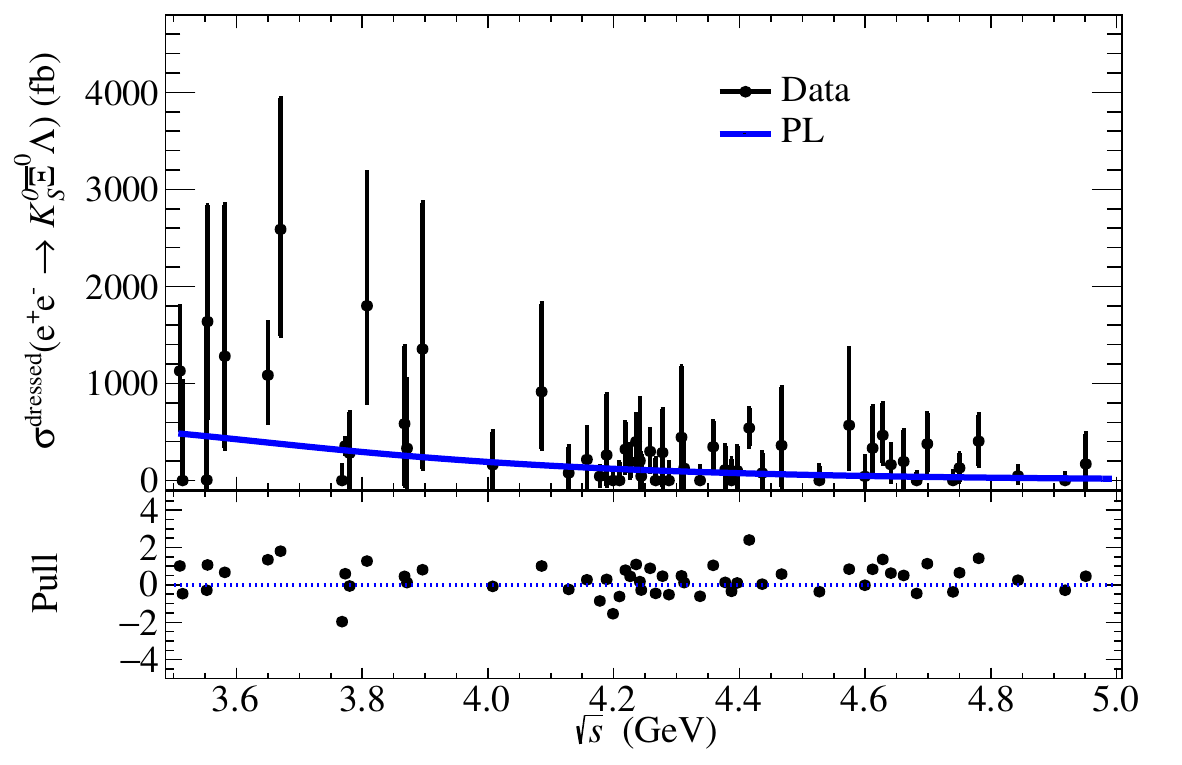}
  \caption{Fits to the dressed cross sections of the reaction $\KXL$ with the assumption of a resonance ($\psi(3770)$, $\psi(4040)$, $\psi(4160)$, $Y(4230)$, $Y(4360)$, $\psi(4415)$, $Y(4500)$, $Y(4660)$ or $Y(4710)$) plus a continuum contribution. The blue solid lines are the fit results.}
  \label{Fig:Fitanother}
\end{figure}

\begin{figure}[H]
  \centering
  \includegraphics[width=0.44\textwidth]{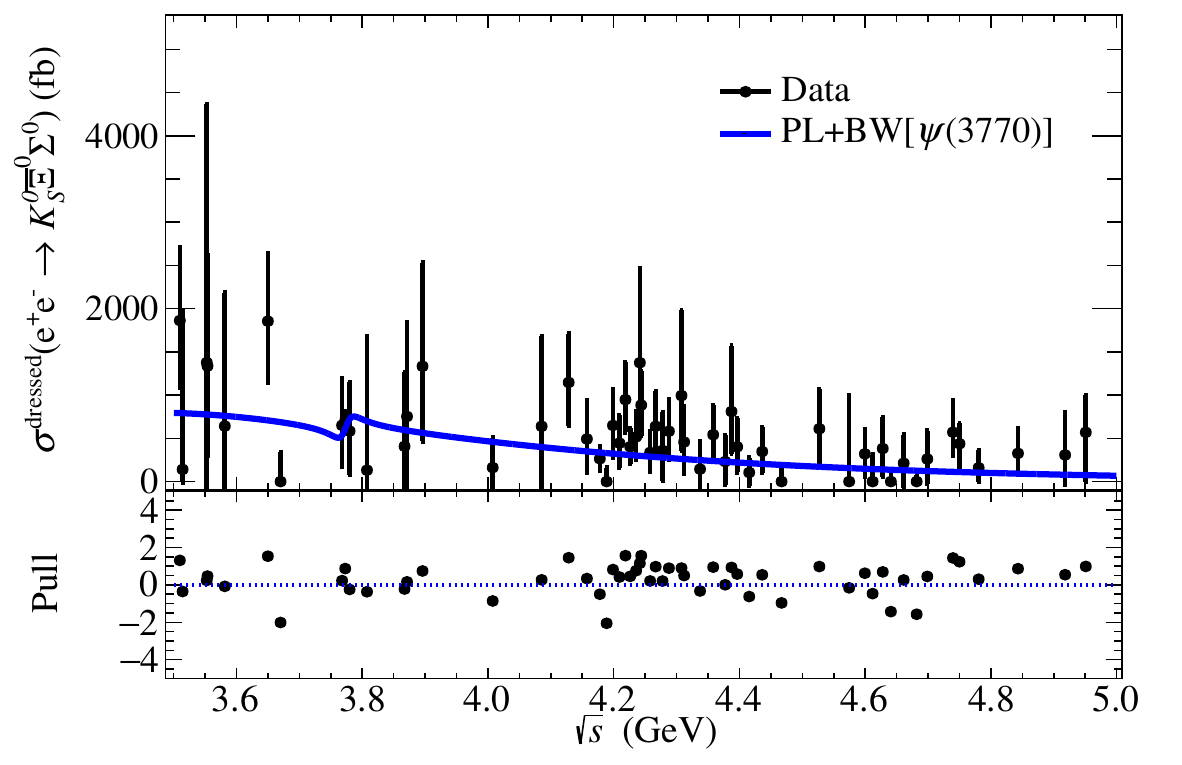}
  \includegraphics[width=0.44\textwidth]{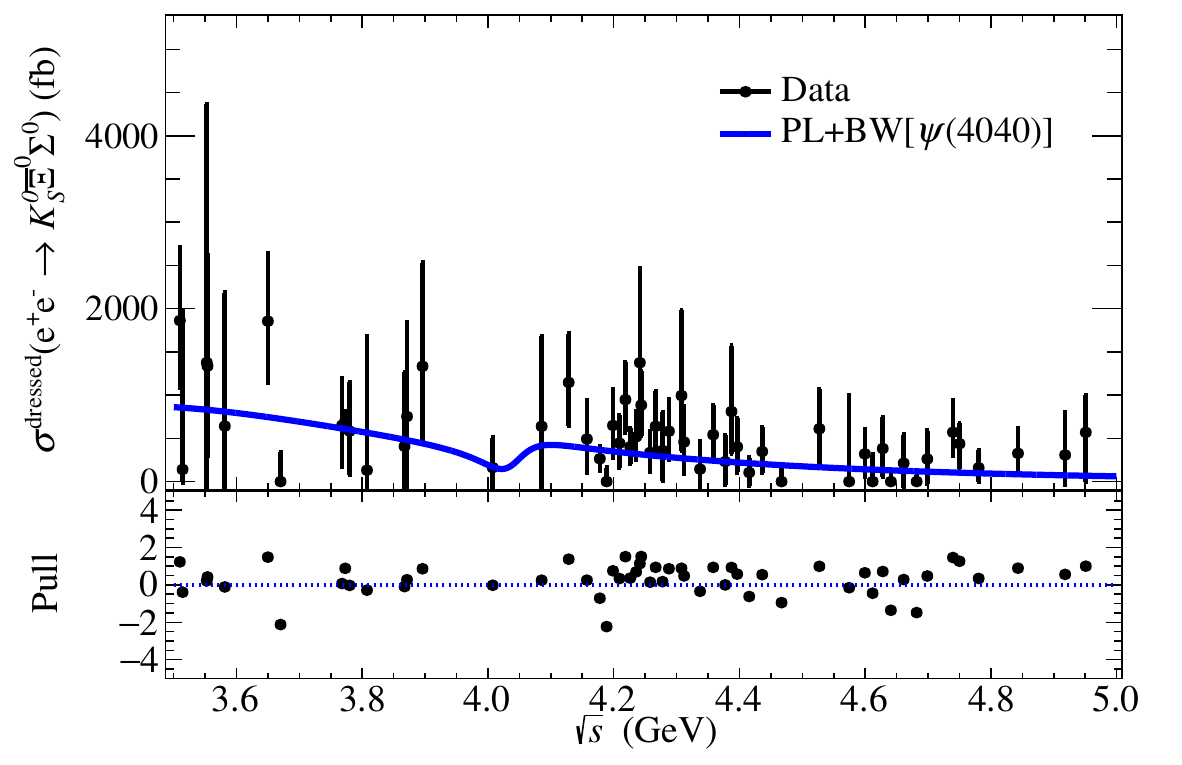}\\
  \includegraphics[width=0.44\textwidth]{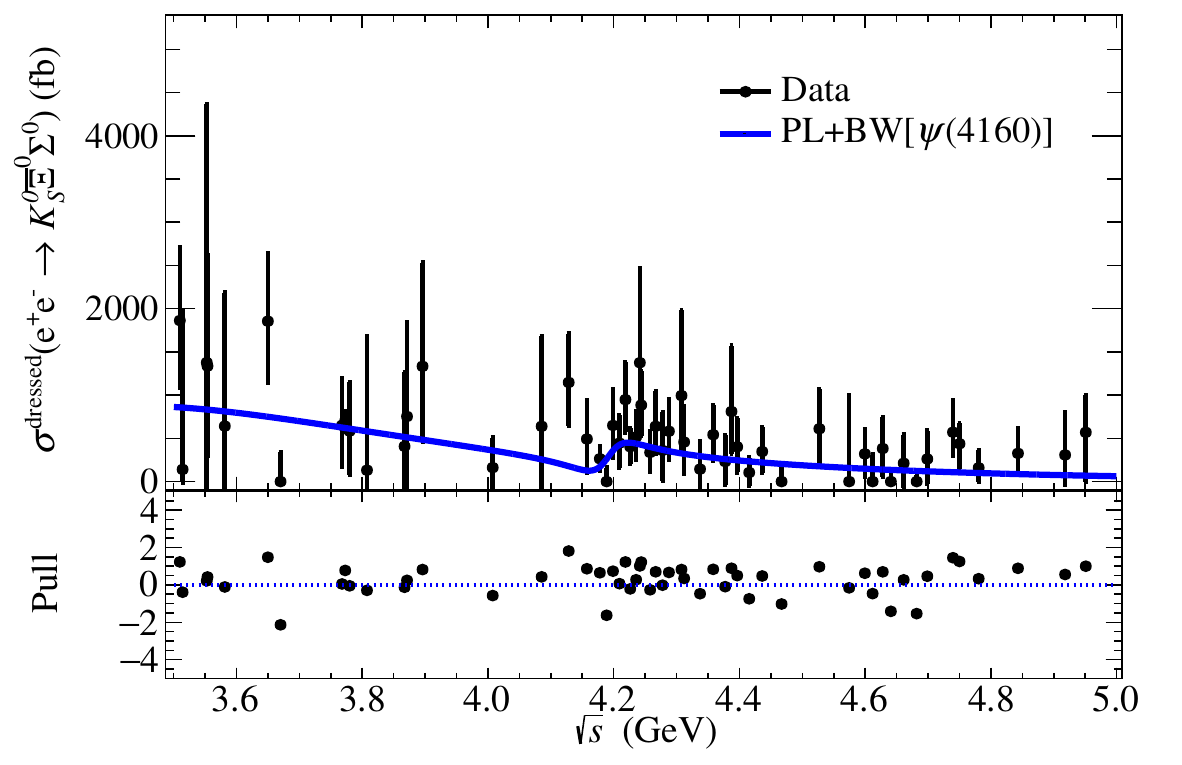}
  \includegraphics[width=0.44\textwidth]{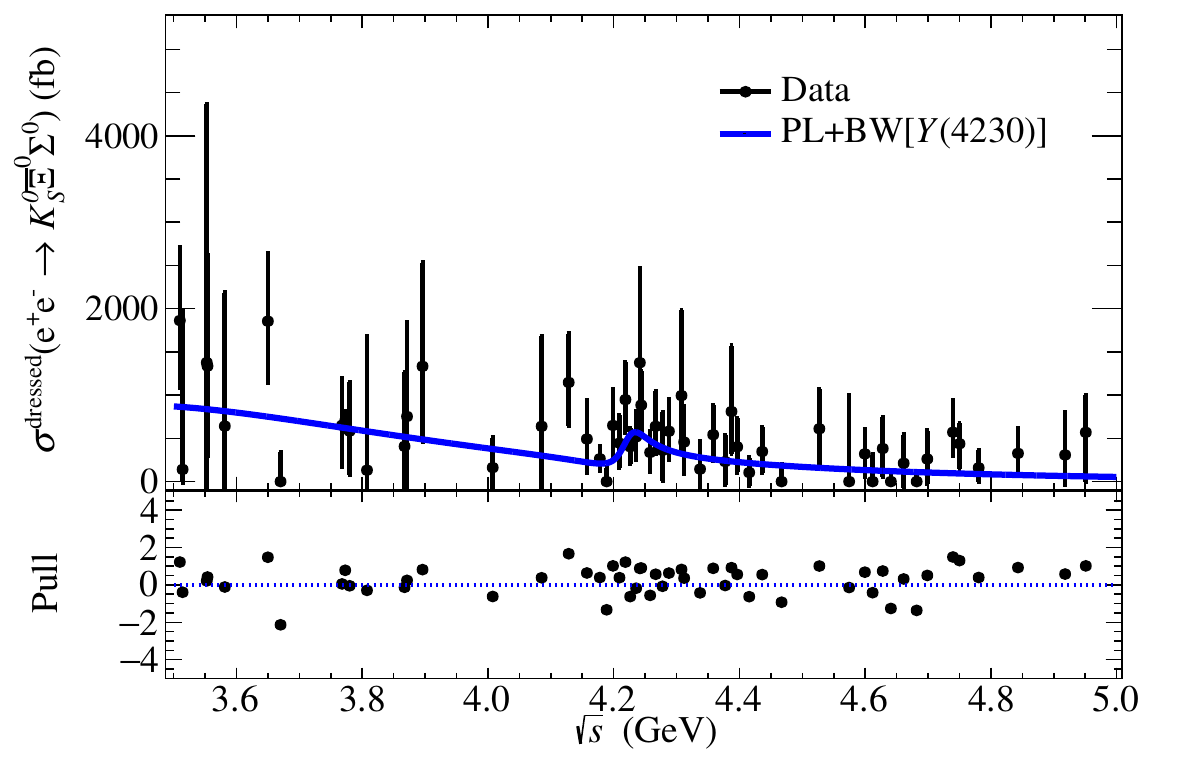}\\
  \includegraphics[width=0.44\textwidth]{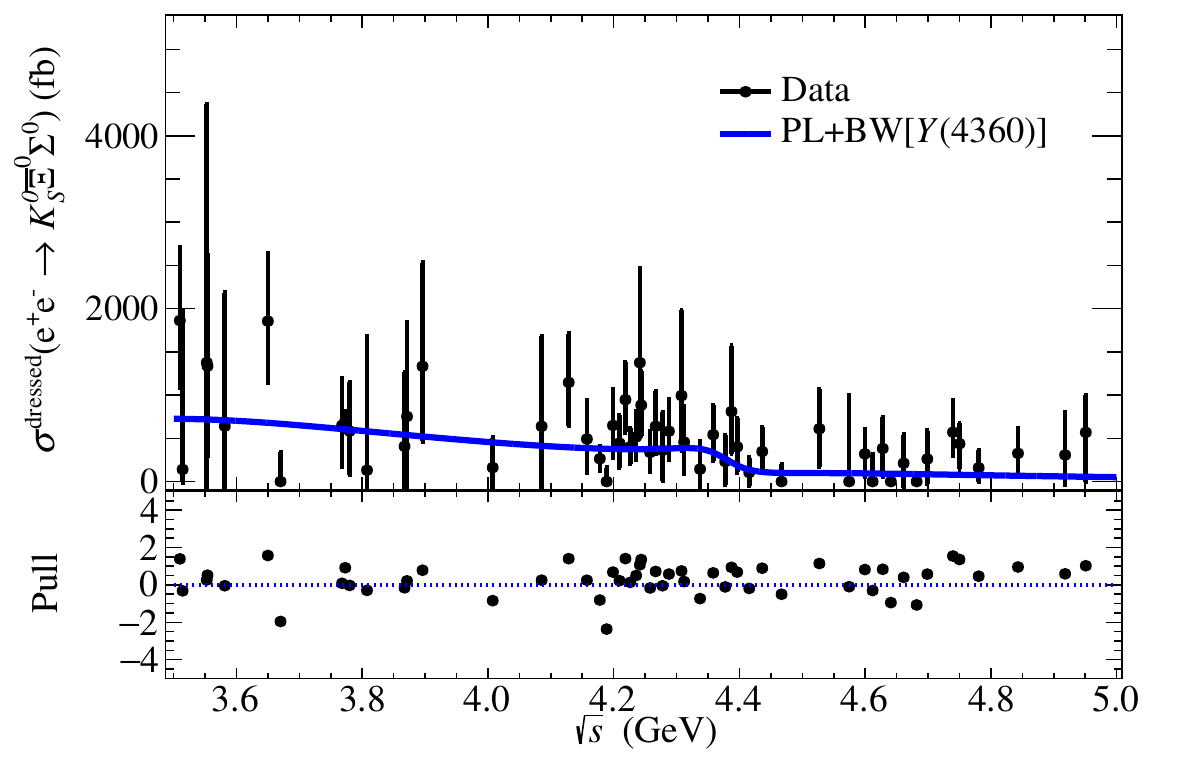}
  \includegraphics[width=0.44\textwidth]{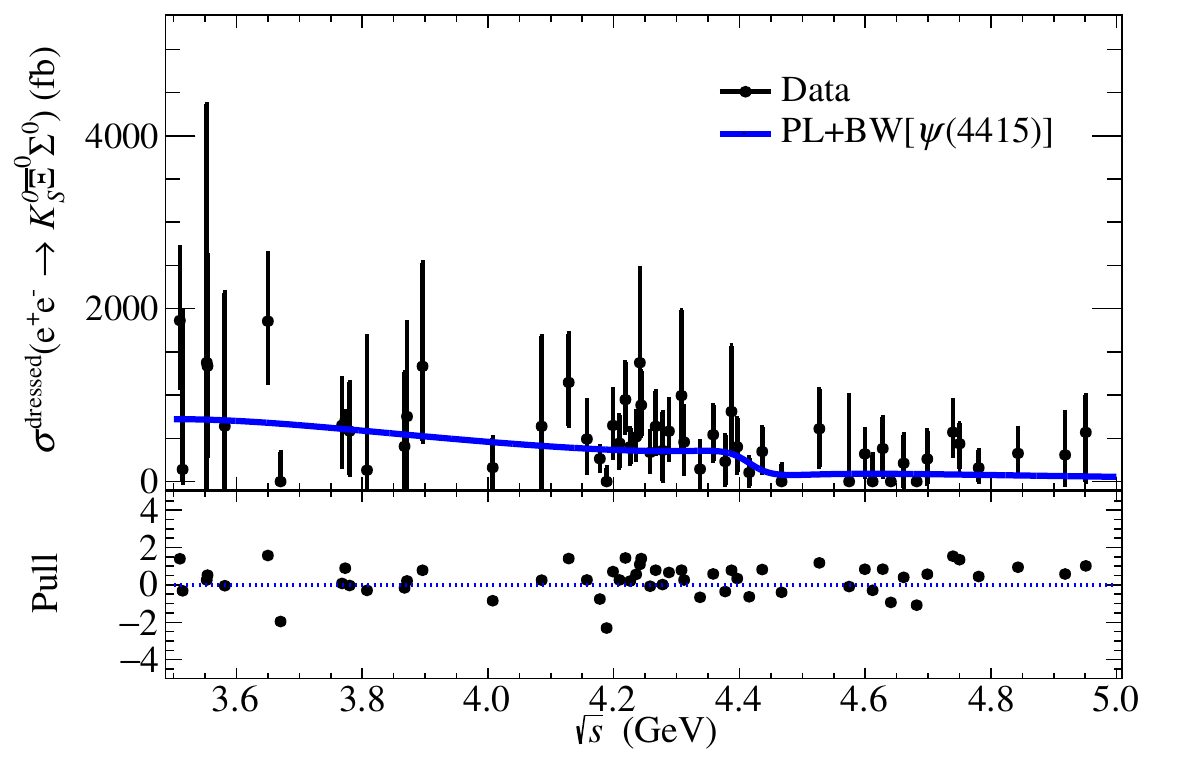}\\
  \includegraphics[width=0.44\textwidth]{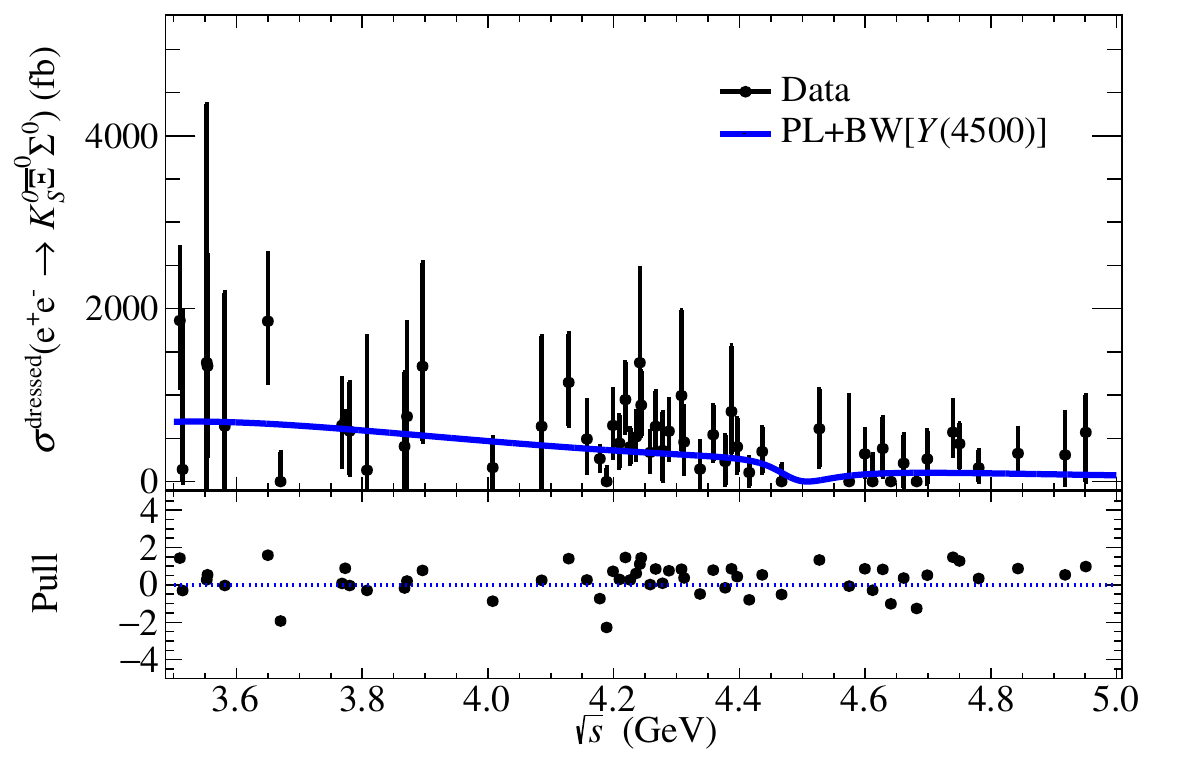}
  \includegraphics[width=0.44\textwidth]{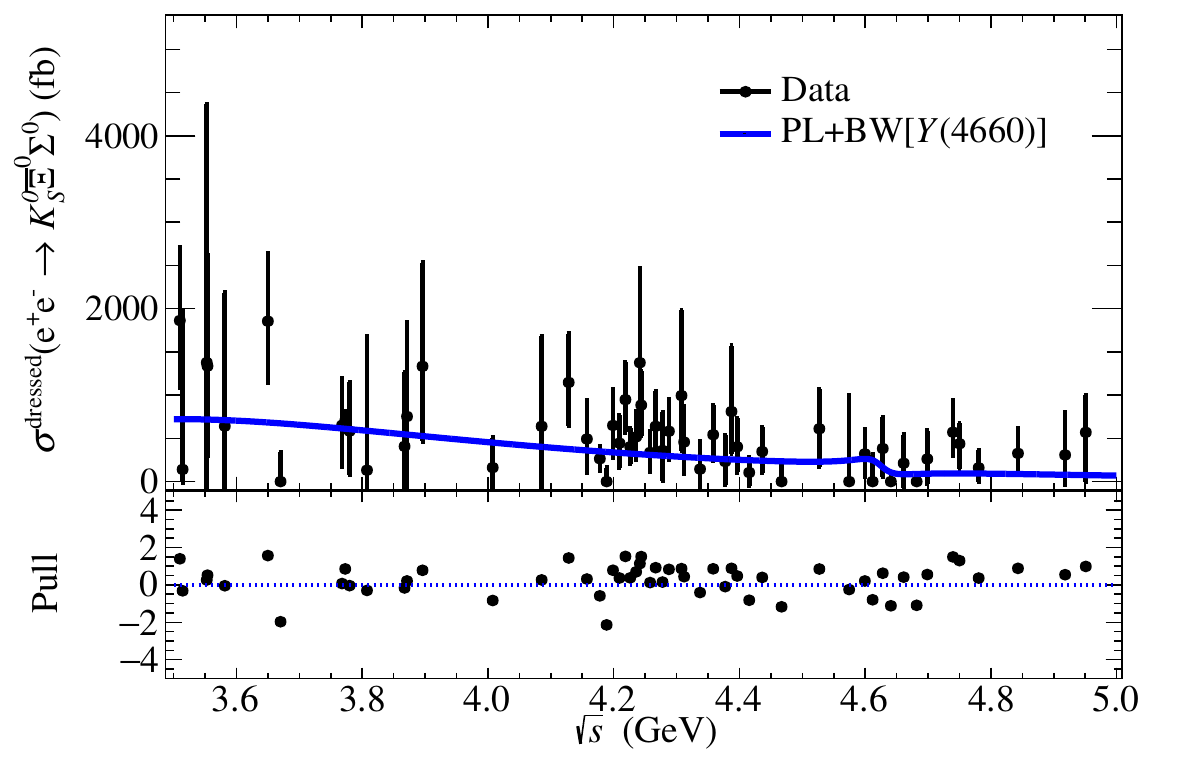}\\
  \includegraphics[width=0.44\textwidth]{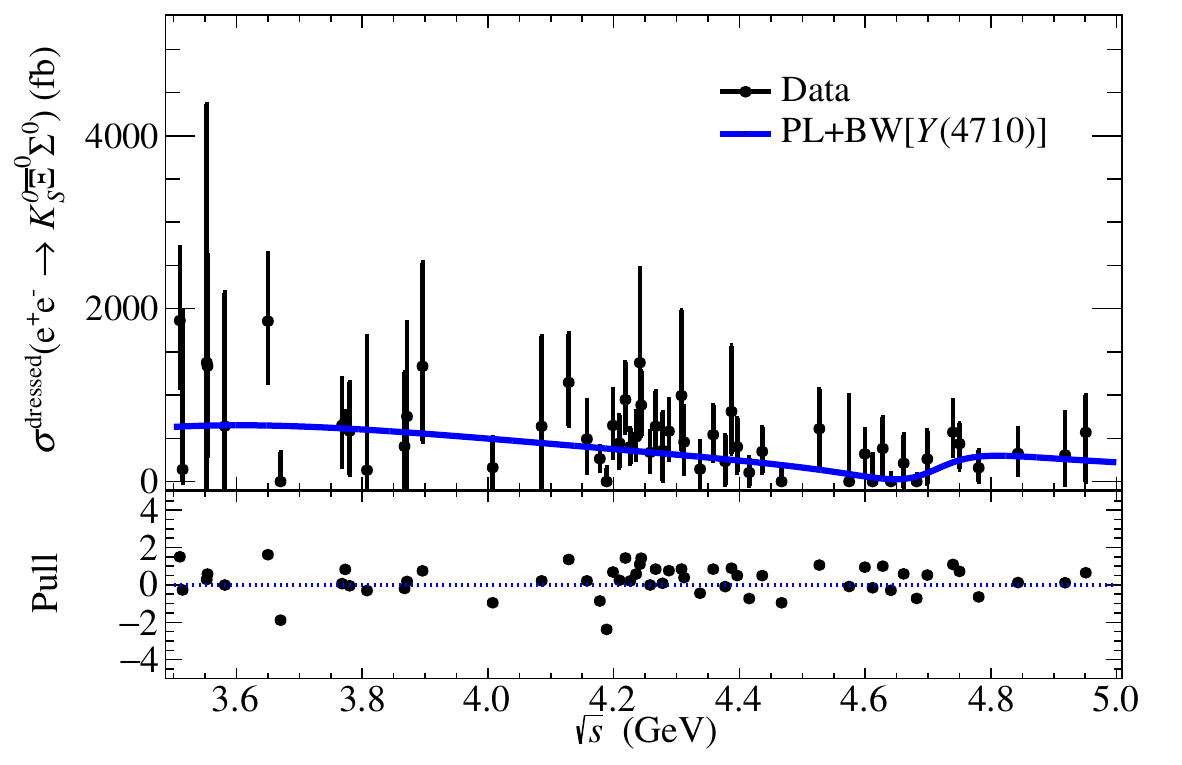}
  \includegraphics[width=0.44\textwidth]{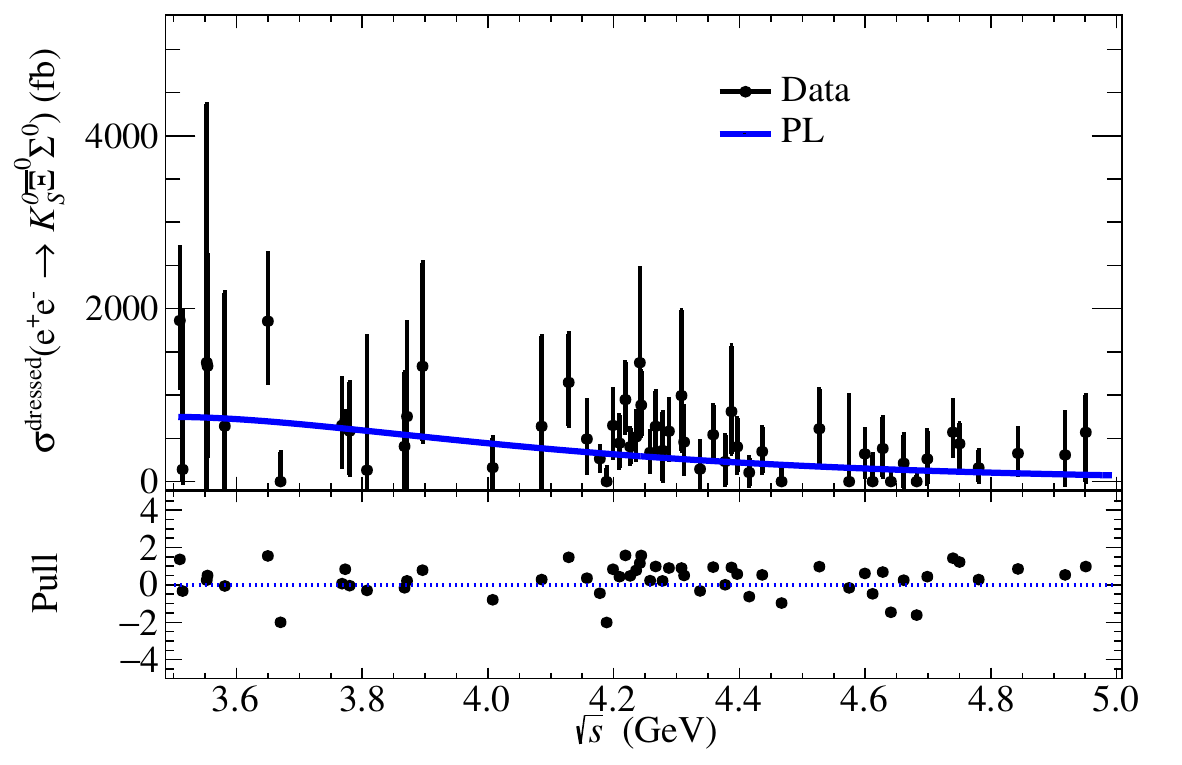}
  \caption{Fits to the dressed cross sections of the reaction $\KXS$ with the assumption of a resonance ($\psi(3770)$, $\psi(4040)$, $\psi(4160)$, $Y(4230)$, $Y(4360)$, $\psi(4415)$, $Y(4500)$, $Y(4660)$ or $Y(4710)$) plus a continuum contribution. The blue solid lines are the fit results.}
  \label{Fig:XiXi::CS::Line-shape_other}
\end{figure}

Assuming a resonance plus a continuum contribution, a fit to the dressed cross sections with the coherent sum of a power-law (PL) function plus a Breit-Wigner (BW) function
\begin{linenomath*}
 \begin{equation}
	\sigma^{\rm dressed}(\sqrt{s}) = \left|{\rm PL}(\sqrt{s}) + e^{i\phi}{\rm BW}(\sqrt{s})\sqrt{\frac{P(\sqrt{s})}{P(M)}}\right|^{2}
\end{equation}
\end{linenomath*}
is applied. Here $\phi$ is the relative phase between the BW function
\begin{linenomath*}
\begin{equation}
   {\rm BW}(\sqrt{s}) = \frac{\sqrt{12\pi\Gamma_{ee}{\cal{B}}\Gamma}}{s-M^{2}+iM\Gamma}
\end{equation}
\end{linenomath*}
and the PL function
\begin{linenomath*}
\begin{equation}
   {\rm PL}(\sqrt{s}) = \frac{c_0\sqrt{P(\sqrt{s})}}{\sqrt{s}^n}.
\end{equation}
\end{linenomath*}
Here, $c_0$ and $n$ are free parameters. When no intermediate resonances are included in the fit, we obtain $(c_0,\ n)=(93.5\pm133.0,\ 8.1\pm1.1)$ for the process $\KXL$ and $(39.2\pm23.0,\ 7.1\pm0.4)$ for the process $\KXS$. The $\sqrt{P(\sqrt{s})}$ is the three-body PHSP factor, the mass $M$ and total width $\Gamma$ are fixed to the PDG values of the assumed resonance~\cite{PDG2020} and the previous BESIII measurement results~\cite{BESIII:cpc1,besiii:y4710}, and $\Gamma_{ee}{\cal{B}}$ is the product of the electronic partial width and the branching fraction for the assumed resonance decaying into the $K_S^0\Xib\Lambda/\Sigma^0$ final state. The significance of a resonance is determined from the change of $\chi^2$ and the number of degrees of freedom $n_{dof}$ for the hypothesis with and without the resonance.
Due to limited statistics, no significant \mbox{charmonium(-like)} states are found in the fit. Thus, the upper limits of the products of branching fractions and the electronic partial widths for these charmonium(-like) states decaying into the $K^0_S\Xib\Lambda/\Sigma^0$ final states, including systematic uncertainties, are provided at the 90\% C.L. using the Bayesian approach~\cite{Zhu:2008ca}. 
Figures~\ref{Fig:Fitanother} and~\ref{Fig:XiXi::CS::Line-shape_other} show the fit to the dressed cross section of $\KXL$ and $\KXS$ with and without resonances included [e.g., $\psi(3770)$, $\psi(4040)$, $\psi(4160)$, $Y(4230)$, $Y(4360)$, $\psi(4415)$, $Y(4500)$, $Y(4660)$ or $Y(4710)$], respectively. The upper limits at the 90\% C.L. on the products of branching fractions and the electronic partial widths for these charmonium(-like) states are listed in table~\ref{tab:multisolution}.

\begin{table}[htbp]
\centering
\caption{\small Fitted parameters to the dressed cross sections for the processes $\KXL$ and $\KXS$ with upper limits on $\Gamma_{ee}\cal B$ or $\cal B$. The fit procedure includes both statistical and systematic uncertainties except for the CM energy calibration. The ${\cal{B}}$ denotes the branching fraction of the assumed resonance decaying into the final state, and ``-'' denotes the absence of a precise measurement for the corresponding excited state in the PDG~\cite{PDG2020}. $\cal{S}$ stands for the signal significance. }
\setlength{\tabcolsep}{12pt}
\begin{tabular}{cccccc}
\hline
\hline
\multicolumn{6}{c}{$\KXL$}\\
\hline
Resonance   	&$M$ ($\text{GeV}/c^2$) 	&$\Gamma$ (GeV) 	&$\Gamma_{ee}\mathcal{B}$ (eV) 	&$\mathcal{B}$ ($\times10^{-4}$) &$\cal{S}~(\sigma)$	\\
\hline
  $\psi(3770)$ 	& 3.7737~\cite{PDG2020} 	& 0.0272~\cite{PDG2020} 	&$\textless$0.05	&$\textless$2.0	&1.9\\
  $\psi(4040)$ 	& 4.0400~\cite{PDG2020} 	& 0.0840~\cite{PDG2020} 	&$\textless$0.15	&$\textless$1.7	&0.2\\
  $\psi(4160)$ 	& 4.1910~\cite{PDG2020} 	& 0.0690~\cite{PDG2020} 	&$\textless$0.05	&$\textless$1.1	&1.5\\
  $Y(4230)$ 	    & 4.2222~\cite{PDG2020} 	& 0.0510~\cite{PDG2020} 	&$\textless$0.05	&-	&1.5\\
  $Y(4360)$ 	    & 4.3740~\cite{PDG2020} 	& 0.1200~\cite{PDG2020} 	&$\textless$0.11	&-	&1.7\\
  $\psi(4415)$ 	& 4.4150~\cite{PDG2020} 	& 0.1100~\cite{PDG2020} 	&$\textless$0.12	&$\textless$3.4	&1.9\\
  $Y(4500)$ 	    & 4.4847~\cite{BESIII:cpc1} 	& 0.1111~\cite{BESIII:cpc1} 	&$\textless$0.09	&-	&1.4\\
  $Y(4660)$ 	    & 4.6230~\cite{PDG2020} 	& 0.0550~\cite{PDG2020} 	&$\textless$0.07	&-	&1.3\\
  $Y(4710)$ 	    & 4.7040~\cite{besiii:y4710} 	 	& 0.1832~\cite{besiii:y4710} 	 	&$\textless$0.09	&-	&0.8\\
\hline
\multicolumn{6}{c}{$\KXS$}\\
\hline
Resonance   	&$M$ ($\text{GeV}/c^2$) 	&$\Gamma$ (GeV) 	&$\Gamma_{ee}\mathcal{B}$ (eV) 	&$\mathcal{B}$ ($\times10^{-4}$) &$\cal{S}~(\sigma)$	\\
\hline
  $\psi(3770)$ 	& 3.7737~\cite{PDG2020} 	& 0.0272~\cite{PDG2020} 	&$\textless$0.08	&$\textless$3.1	&0.1\\
  $\psi(4040)$ 	& 4.0400~\cite{PDG2020} 	& 0.0840~\cite{PDG2020} 	&$\textless$0.25	&$\textless$2.9	&0.6\\
  $\psi(4160)$ 	& 4.1910~\cite{PDG2020} 	& 0.0690~\cite{PDG2020} 	&$\textless$0.12	&$\textless$2.5	&1.6\\
  $Y(4230)$ 	    & 4.2222~\cite{PDG2020} 	& 0.0510~\cite{PDG2020} 	&$\textless$0.11	&-	&2.1\\
  $Y(4360)$ 	    & 4.3740~\cite{PDG2020} 	& 0.1200~\cite{PDG2020} 	&$\textless$0.21	&- &1.3\\
  $\psi(4415)$ 	& 4.4150~\cite{PDG2020} 	& 0.1100~\cite{PDG2020} 	&$\textless$0.17	&$\textless$4.7	&1.2\\
  $Y(4500)$ 	    & 4.4847~\cite{BESIII:cpc1} 	& 0.1111~\cite{BESIII:cpc1} 	&$\textless$0.15	&-	&0.9\\
  $Y(4660)$ 	    & 4.6230~\cite{PDG2020} 	& 0.0550~\cite{PDG2020} 	&$\textless$0.07	&-	&1.3\\
  $Y(4710)$ 	    & 4.7040~\cite{besiii:y4710} 	& 0.1832~\cite{besiii:y4710} 	 	&$\textless$0.24	&-	&1.9\\
\hline
\hline 
\end{tabular}
\label{tab:multisolution}
\end{table}

 \section{Summary}
Using a data sample of $\EE$ collisions corresponding to a total integrated luminosity of \SI{44.55}{fb^{-1}} collected with the BESIII detector at the BEPCII collider, we measured the Born cross sections for the processes $\KXX$ at 56 center-of-mass energies between 3.510 and \SI{4.951}{GeV} with a partial reconstruction technique. A fit to the dressed cross sections for the reactions $\KXX$ under the assumption of one resonance, e.g., $\psi(3770)$, $\psi(4040)$, $\psi(4160)$, $Y(4230)$, $Y(4360)$, $\psi(4415)$, $Y(4500)$, $Y(4660)$ or $Y(4710)$ plus a continuum contribution is performed. The fitted parameters, $\Gamma_{ee}{\cal{B}}$, for each assumed resonance are summarized in table~\ref{tab:multisolution}. No statistically significant evidence for any of the considered resonances decaying into the $K_S^0\Xib\Lambda/\Sigma^0$ final states is found. The upper limits for the products of the electronic partial widths and branching fractions for each assumed resonance decaying into the $K_S^0\Xib\Lambda/\Sigma^0$ final state are set at the $90\%$ C.L. 
These results provide the first experimental constraints on the processes $e^+e^- \to K_S^0\bar{\Xi}^0\Lambda/\Sigma^0$ and may shed light on the nature of baryonic production above the open-charm region.

\acknowledgments
\noindent
The BESIII Collaboration thanks the staff of BEPCII (https://cstr.cn/31109.02.BEPC) and the IHEP computing center for their strong support. This work is supported in part by National Key R\&D Program of China under Contracts Nos. 2023YFA1606000, 2023YFA1606704, 2025YFA1613900; National Natural Science Foundation of China (NSFC) under Contracts Nos. 
12247101,
11635010, 11935015, 11935016, 11935018, 12025502, 12035009, 12035013, 12061131003, 12192260, 12192261, 12192262, 12192263, 12192264, 12192265, 12221005, 12225509, 12235017, 12342502, 12361141819, 12535005; 
the Fundamental Research Funds for the Central Universities No.
lzujbky-2025-ytA05, No. lzujbky-2025-it06, No. lzujbky-2024-jdzx06;
the Natural Science Foundation of Gansu Province No. 22JR5RA389, No. 25JRRA799;
the "111 Center" under Grant No. B20063;
the Chinese Academy of Sciences (CAS) Large-Scale Scientific Facility Program; the Strategic Priority Research Program of Chinese Academy of Sciences under Contract No. XDA0480600; CAS under Contract No. YSBR-101; 100 Talents Program of CAS; The Institute of Nuclear and Particle Physics (INPAC) and Shanghai Key Laboratory for Particle Physics and Cosmology; Agencia Nacional de Investigación y Desarrollo de Chile (ANID), Chile under Contract No. ANID CCTVal CIA250027; ERC under Contract No. 758462; German Research Foundation DFG under Contract No. FOR5327; Istituto Nazionale di Fisica Nucleare, Italy; Knut and Alice Wallenberg Foundation under Contracts Nos. 2021.0174, 2021.0299, 2023.0315; Ministry of Development of Turkey under Contract No. DPT2006K-120470; National Research Foundation of Korea under Contract No. RS-2026-25486791; National Science and Technology fund of Mongolia; Polish National Science Centre under Contract No. 2024/53/B/ST2/00975; STFC (United Kingdom); Swedish Research Council under Contract No. 2019.04595; U. S. Department of Energy under Contract No. DE-FG02-05ER41374.

\newpage
{\bf \noindent The BESIII collaboration}\\
\\
{\small
M.~Ablikim$^{1}$\BESIIIorcid{0000-0002-3935-619X},
M.~N.~Achasov$^{4,c}$\BESIIIorcid{0000-0002-9400-8622},
P.~Adlarson$^{84}$\BESIIIorcid{0000-0001-6280-3851},
X.~C.~Ai$^{90}$\BESIIIorcid{0000-0003-3856-2415},
C.~S.~Akondi$^{32A,32B}$\BESIIIorcid{0000-0001-6303-5217},
R.~Aliberti$^{40}$\BESIIIorcid{0000-0003-3500-4012},
A.~Amoroso$^{83A,83C}$\BESIIIorcid{0000-0002-3095-8610},
Q.~An$^{79,66,\dagger}$,
Y.~H.~An$^{90}$\BESIIIorcid{0009-0008-3419-0849},
M.~S.~Anderson$^{40}$\BESIIIorcid{0009-0008-1550-2632}
% ...
\\
\vspace{0.2cm}
(BESIII Collaboration)\\
\vspace{0.2cm} {\it
$^{1}$ Institute of High Energy Physics, Beijing 100049, People's Republic of China\\
% ...
$^{4}$ Budker Institute of Nuclear Physics SB RAS (BINP), Novosibirsk 630090, Russia\\
% ...
$^{32}$ INFN Sezione di Ferrara, (A)INFN Sezione di Ferrara, I-44122, Ferrara, Italy; (B)University of Ferrara, I-44122, Ferrara, Italy\\
% ...
$^{40}$ Johannes Gutenberg University of Mainz, Johann-Joachim-Becher-Weg 45, D-55099 Mainz, Germany\\
% ...
$^{66}$ State Key Laboratory of Particle Detection and Electronics, Beijing 100049, Hefei 230026, People's Republic of China\\
% ...
$^{79}$ University of Science and Technology of China, Hefei 230026, People's Republic of China\\
% ...
$^{83}$ University of Turin and INFN, (A)University of Turin, I-10125, Turin, Italy; (B)University of Eastern Piedmont, I-15121, Alessandria, Italy; (C)INFN, I-10125, Turin, Italy\\
$^{84}$ Uppsala University, Box 516, SE-75120 Uppsala, Sweden\\
% ...
$^{90}$ Zhengzhou University, Zhengzhou 450001, People's Republic of China\\

\vspace{0.2cm}
$^{\dagger}$ Deceased\\
% ...
$^{c}$ Also at the Novosibirsk State University, Novosibirsk, 630090, Russia\\
% ...
}
}


\begin{thebibliography}{99}

\bibitem{Brambilla:2010cs}
N.~Brambilla, \textit{et al.},
\textit{Heavy quarkonium: progress, puzzles, and opportunities},
\textcolor{blue}{\href{https://doi.org/10.1140/epjc/s10052-010-1534-9} {\textit{Eur. Phys. J. C} \textbf{71} (2011) 1534}}
[\textcolor{blue}{\href{https://arxiv.org/abs/1010.5827}{arXiv:hep-ph/1010.5827}}] 
[\textcolor{blue}{\href{https://inspirehep.net/literature/874793}{\textsc{inSPIRE}}}].

\bibitem{Briceno:2015rlt}
R.~A.~Briceno, \textit{et al.},
\textit{Issues and opportunities in exotic hadrons},
\textcolor{blue}{\href{https://doi.org/10.1088/1674-1137/40/4/042001} {\textit{Chin. Phys. C} \textbf{40} (2016) 042001}}
[\textcolor{blue}{\href{https://arxiv.org/abs/1511.06779}{arXiv:1511.06779}}] 
[\textcolor{blue}{\href{https://inspirehep.net/literature/1405969}{\textsc{inSPIRE}}}].

\bibitem{PR_2020}
N.~Brambilla \textit{et al.},
\textit{The XYZ states: Experimental and theoretical status and perspectives},
\textcolor{blue}{\href{https://doi.org/10.1016/j.physrep.2020.05.001} {\textit{Phys. Rept.} \textbf{873} (2020) 1}}
[\textcolor{blue}{\href{https://arxiv.org/abs/1907.07583}{arXiv:1907.07583}}] 
[\textcolor{blue}{\href{https://inspirehep.net/literature/1744286}{\textsc{inSPIRE}}}].

\bibitem{XYZ:states}
G.~Mezzadri and S.~Spataro,
\textit{XYZ states: An experimental point-of-view},
\textcolor{blue}{\href{https://doi.org/10.1016/j.revip.2022.100070} {\textit{Rev. Phys.} \textbf{8} (2022) 100070}}
[\textcolor{blue}{\href{https://inspirehep.net/literature/2065863}{\textsc{inSPIRE}}}].

\bibitem{Recent_XYZ}
R.~A.~Briceno, \textit{et al.},
\textit{Recent XYZ studies at BESIII},
\textcolor{blue}{\href{https://doi.org/10.1016/j.nuclphysbps.2022.09.014} {\textit{Nucl. Part. Phys. Proc.} \textbf{318} (2022) 61}}
[\textcolor{blue}{\href{https://inspirehep.net/literature/2513313}{\textsc{inSPIRE}}}].

\bibitem{XYZ_RMP98}
X.~Wang, X.~Liu, and Y.~Gao, 
\textit{Colloquium: Hadron production
in open-charm meson pairs at e+e- colliders},
\textcolor{blue}{\href{https://doi.org/10.1103/2mrp-chly} {\textit{ Rev. Mod. Phys.} \textbf{98} (2026) 021001}}
[\textcolor{blue}{\href{https://arxiv.org/abs/2502.15117}{arXiv:2502.15117}}] 
[\textcolor{blue}{\href{https://inspirehep.net/literature/2893289}{\textsc{inSPIRE}}}].

\bibitem{Barnes:2005pb}
T.~Barnes, S.~Godfrey and E.~S.~Swanson,
\textit{Higher charmonia},
\textcolor{blue}{\href{https://doi.org/10.1103/PhysRevD.72.054026} {\textit{Phys. Rev. D} \textbf{72} (2005) 054026}}
[\textcolor{blue}{\href{https://arxiv.org/abs/hep-ph/0505002}{arXiv:hep-ph/0505002}}] 
[\textcolor{blue}{\href{https://inspirehep.net/literature/681760}{\textsc{inSPIRE}}}].

\bibitem{BES:2001ckj}
BES Collaboration,
\textit{Measurements of the cross section for $e^+ e^- \to hadrons$ at center-of-mass energies from 2 GeV to 5 GeV},
\textcolor{blue}{\href{https://doi.org/10.1103/PhysRevLett.88.101802} {\textit{Phys. Rev. Lett.} \textbf{88} (2002) 101802}}
[\textcolor{blue}{\href{https://arxiv.org/abs/hep-ex/0102003}{arXiv:hep-ex/0102003}}] 
[\textcolor{blue}{\href{https://inspirehep.net/literature/552757}{\textsc{inSPIRE}}}].


\bibitem{RVUE:2005}
K.~K.~Seth,
\textit{Alternative analysis of the R measurements: Resonance parameters of the higher vector states of charmonium},
\textcolor{blue}{\href{https://doi.org/10.1103/PhysRevD.72.017501} {\textit{Phys. Rev. Lett.} \textbf{72} (2005) 017501}}
[\textcolor{blue}{\href{https://arxiv.org/abs/hep-ex/0405007}{arXiv:hep-ex/0405007}}] 
[\textcolor{blue}{\href{https://inspirehep.net/literature/649689}{\textsc{inSPIRE}}}].


\bibitem{BES:2008}
BES Collaboration,
\textit{Determination of the $\psi(3770)$, $\psi(4040)$, $\psi(4160)$ and $\psi(4415)$ resonance parameters},
\textcolor{blue}{\href{https://doi.org/10.1016/j.physletb.2007.11.100} {\textit{Phys. Lett. B} \textbf{660} (2008) 315}}
[\textcolor{blue}{\href{https://arxiv.org/abs/0705.4500}{arXiv:0705.4500}}] 
[\textcolor{blue}{\href{https://inspirehep.net/literature/751833}{\textsc{inSPIRE}}}].


\bibitem{RVUE:2010}
X.~H.~Mo, C.~Z.~Yuan, and P.~Wang,
\textit{On the leptonic partial widths of the excited $\psi$ states},
\textcolor{blue}{\href{https://doi.org/10.1103/PhysRevD.82.077501} {\textit{Phys. Rev. D} \textbf{82} (2010) 077501}}
[\textcolor{blue}{\href{https://arxiv.org/abs/1007.0084}{arXiv:1007.0084}}] 
[\textcolor{blue}{\href{https://inspirehep.net/literature/860039}{\textsc{inSPIRE}}}].


\bibitem{D0D0_2024}
BESIII Collaboration,
\textit{Precise Measurement of Born Cross Sections for ${e}^{+}{e}^{\ensuremath{-}}\ensuremath{\rightarrow}D\overline{D}$ at $\sqrt{s}=3.80\ensuremath{-}4.95\text{ }\text{ }\mathrm{GeV}$},
\textcolor{blue}{\href{https://doi.org/10.1103/PhysRevLett.133.081901} {\textit{Phys. Rev. Lett.} \textbf{133} (2024) 081901}}
[\textcolor{blue}{\href{https://arxiv.org/abs/2402.03829}{arXiv:2402.03829}}] 
[\textcolor{blue}{\href{https://inspirehep.net/literature/2755997}{\textsc{inSPIRE}}}].


\bibitem{BaBar:2005hhc}
\textsc{BaBar} Collaboration,
\textit{Observation of a broad structure in the $\pi^+ \pi^- J/\psi$ mass spectrum around 4.26 GeV/c$^2$},
\textcolor{blue}{\href{https://doi.org/10.1103/PhysRevLett.95.142001} {\textit{Phys. Rev. Lett.} \textbf{95} (2005) 142001}}
[\textcolor{blue}{\href{https://arxiv.org/abs/hep-ex/0506081}{arXiv:hep-ex/0506081}}] 
[\textcolor{blue}{\href{https://inspirehep.net/literature/686354}{\textsc{inSPIRE}}}].

\bibitem{BaBar:2006ait}
\textsc{BaBar} Collaboration,
\textit{Evidence of a broad structure at an invariant mass of 4.32 $GeV/c^{2}$ in the reaction $e^{+} e^{-} \to \pi^{+} \pi^{-} \psi(2S)$ measured at \textsc{BaBar}},
\textcolor{blue}{\href{https://doi.org/10.1103/PhysRevLett.98.212001} {\textit{Phys. Rev. Lett.} \textbf{98} (2007) 212001}}
[\textcolor{blue}{\href{https://arxiv.org/abs/hep-ex/0610057}{arXiv:hep-ex/0610057}}] 
[\textcolor{blue}{\href{https://inspirehep.net/literature/729388}{\textsc{inSPIRE}}}].

\bibitem{Belle:2007umv}
Belle Collaboration,
\textit{Observation of two resonant structures in ${e}^{+}{e}^{\ensuremath{-}}\ensuremath{\rightarrow}{\ensuremath{\pi}}^{+}{\ensuremath{\pi}}^{\ensuremath{-}}\ensuremath{\psi}(2S)$ via initial-state radiation at Belle},
\textcolor{blue}{\href{https://doi.org/10.1103/PhysRevLett.99.142002} {\textit{Phys. Rev. Lett.} \textbf{99} (2007) 142002}}
[\textcolor{blue}{\href{https://arxiv.org/abs/0707.3699}{arXiv:0707.3699}}] 
[\textcolor{blue}{\href{https://inspirehep.net/literature/756643}{\textsc{inSPIRE}}}].

\bibitem{Belle:2007dxy}
Belle Collaboration,
\textit{Measurement of $e^+e^-\to \pi^+\pi^- J/\psi$ cross section via initial state radiation at Belle},
\textcolor{blue}{\href{https://doi.org/10.1103/PhysRevLett.99.182004} {\textit{Phys. Rev. Lett.} \textbf{99} (2007) 182004}}
[\textcolor{blue}{\href{https://arxiv.org/abs/0707.2541}{arXiv:0707.2541}}] 
[\textcolor{blue}{\href{https://inspirehep.net/literature/756012}{\textsc{inSPIRE}}}].


\bibitem{BaBar:2012vyb}
\textsc{BaBar} Collaboration,
\textit{Study of the reaction $e^{+}e^{-} \to J/\psi\pi^{+}\pi^{-}$ via initial-state radiation at \textsc{BaBar}},
\textcolor{blue}{\href{https://doi.org/10.1103/PhysRevD.86.051102} {\textit{Phys. Rev. D} \textbf{86} (2012) 051102}}
[\textcolor{blue}{\href{https://arxiv.org/abs/1204.2158}{arXiv:1204.2158}}] 
[\textcolor{blue}{\href{https://inspirehep.net/literature/1107905}{\textsc{inSPIRE}}}].

\bibitem{Belle:2013yex}
Belle Collaboration,
\textit{Study of ${e}^{\mathbf{+}}{e}^{\mathbf{\ensuremath{-}}}\ensuremath{\rightarrow}{\ensuremath{\pi}}^{\mathbf{+}}{\ensuremath{\pi}}^{\mathbf{\ensuremath{-}}}J/\ensuremath{\psi}$ and observation of a charged charmoniumlike state at Belle},
\textcolor{blue}
{\href{https://doi.org/10.1103/PhysRevLett.110.252002} {\textit{Phys. Rev. Lett.} \textbf{110} (2013) 252002}}{\href{https://doi.org/10.1103/PhysRevLett.111.019901} {[erratum: \textit{Phys. Rev. Lett.} \textbf{111} (2013) 019901]}}
[\textcolor{blue}{\href{https://arxiv.org/abs/1304.0121}{arXiv:1304.0121}}] 
[\textcolor{blue}{\href{https://inspirehep.net/literature/1225975}{\textsc{inSPIRE}}}].

\bibitem{BaBar:2012hpr}
\textsc{BaBar} Collaboration,
\textit{Study of the reaction $e^{+}e^{-}\to \psi(2S)\pi^{+}\pi^{-}$ via initial-state radiation at \textsc{BaBar}},
\textcolor{blue}{\href{https://doi.org/10.1103/PhysRevD.89.111103} {\textit{Phys. Rev. D} \textbf{89} (2014) 111103}}
[\textcolor{blue}{\href{https://arxiv.org/abs/1211.6271}{arXiv:1211.6271}}] 
[\textcolor{blue}{\href{https://inspirehep.net/literature/1204444}{\textsc{inSPIRE}}}].

\bibitem{Belle:2014wyt}
Belle Collaboration,
\textit{Measurement of $e^+e^- \to \pi^+\pi^-\psi(2S)$ via Initial State Radiation at Belle},
\textcolor{blue}{\href{https://doi.org/10.1103/PhysRevD.91.112007} {\textit{Phys. Rev. D} \textbf{91} (2015) 112007}}
[\textcolor{blue}{\href{https://arxiv.org/abs/1410.7641}{arXiv:1410.7641}}] 
[\textcolor{blue}{\href{https://inspirehep.net/literature/1324785}{\textsc{inSPIRE}}}].

\bbt{CLEO}
CLEO Collaboration,
\textit{Charmonium decays of $Y(4260)$, $\psi(4160)$ and $\psi(4040)$},
\textcolor{blue}{\href{https://doi.org/10.1103/PhysRevLett.96.162003} {\textit{Phys. Rev. Lett.} \textbf{96} (2006) 162003}}
[\textcolor{blue}{\href{https://arxiv.org/abs/hep-ex/0602034}{arXiv:hep-ex/0602034}}] 
[\textcolor{blue}{\href{https://inspirehep.net/literature/710864}{\textsc{inSPIRE}}}].

\bbt{BESIIIAB}
BESIII Collaboration,
\textit{Study of $e^+e^-\to\omega\chi_{cJ}$ at center-of-mass energies from 4.21 to 4.42 GeV},
\textcolor{blue}{\href{https://doi.org/10.1103/PhysRevLett.114.092003} {\textit{Phys. Rev. Lett.} \textbf{114} (2015) 092003}}
[\textcolor{blue}{\href{https://arxiv.org/abs/1410.6538}{arXiv:1410.6538}}] 
[\textcolor{blue}{\href{https://inspirehep.net/literature/1323621}{\textsc{inSPIRE}}}].

\bbt{BESIII:cpc1}
BESIII Collaboration,
\textit{Observation of the $Y(4230)$ and a new structure in $\boldsymbol e^+\boldsymbol e^- \boldsymbol\rightarrow \boldsymbol K^+\boldsymbol K^-\boldsymbol J/\boldsymbol\psi$},
\textcolor{blue}{\href{https://doi.org/10.1088/1674-1137/ac945c} {\textit{Chin. Phys. C.} \textbf{46} (2022) 111002}}
[\textcolor{blue}{\href{https://arxiv.org/abs/2204.07800}{arXiv:2204.07800}}] 
[\textcolor{blue}{\href{https://inspirehep.net/literature/2068180}{\textsc{inSPIRE}}}].

\bibitem{BESIII:2023cmv}
BESIII Collaboration,
\textit{Observation of three charmonium-like states with ${J}^{PC}={1}^{\ensuremath{-}\ensuremath{-}}$ in ${e}^{+}{e}^{\ensuremath{-}}\ensuremath{\rightarrow}{D}^{*0}{D}^{*\ensuremath{-}}{\ensuremath{\pi}}^{+}$},
\textcolor{blue}{\href{https://doi.org/10.1103/PhysRevLett.130.121901} {\textit{Phys. Rev. Lett.} \textbf{130} (2023) 121901}}
[\textcolor{blue}{\href{https://arxiv.org/abs/2301.07321}{arXiv:2301.07321}}] 
[\textcolor{blue}{\href{https://inspirehep.net/literature/2645388}{\textsc{inSPIRE}}}].

\bibitem{BESIII:2023cmv1}
BESIII Collaboration,
\textit{Observation of a vector charmoniumlike state at $4.7\text{ }\text{ }\mathrm{GeV}/{c}^{2}$ and search for ${Z}_{cs}$ in ${e}^{+}{e}^{\ensuremath{-}}\ensuremath{\rightarrow}{K}^{+}{K}^{\ensuremath{-}}J/\ensuremath{\psi}$},
\textcolor{blue}{\href{https://doi.org/10.1103/PhysRevLett.131.211902} {\textit{Phys. Rev. Lett.} \textbf{131} (2023) 211902}}
[\textcolor{blue}{\href{https://arxiv.org/abs/2308.15362}{arXiv:2308.15362}}] 
[\textcolor{blue}{\href{https://inspirehep.net/literature/2691894}{\textsc{inSPIRE}}}].

\bibitem{Close:2005iz}
F.~E.~Close and P.~R.~Page,
\textit{Gluonic charmonium resonances at \textsc{BaBar} and BELLE?},
\textcolor{blue}{\href{https://doi.org/10.1016/j.physletb.2005.09.016} {\textit{Phys. Lett. B} \textbf{628} (2005) 215-222}}
[\textcolor{blue}{\href{https://arxiv.org/abs/hep-ph/0507199}{arXiv:hep-ph/0507199}}] 
[\textcolor{blue}{\href{https://inspirehep.net/literature/687628}{\textsc{inSPIRE}}}].



\bibitem{Chen:2016qju}
H.~X.~Chen, W.~Chen, X.~Liu and S.~L.~Zhu,
\textit{The hidden-charm pentaquark and tetraquark states},
\textcolor{blue}{\href{https://doi.org/10.1016/j.physrep.2016.05.004} {\textit{Phys. Rept.} \textbf{639} (2016) 1-121}}
[\textcolor{blue}{\href{https://arxiv.org/abs/1601.02092}{arXiv:1601.02092}}] 
[\textcolor{blue}{\href{https://inspirehep.net/literature/1414795}{\textsc{inSPIRE}}}].

\bibitem{Wang:2019mhs}
J.~Z.~Wang, D.Y.~Chen, X.~Liu and T.~Matsuki,
\textit{Constructing $J/\psi$ family with updated data of charmoniumlike $Y$ states},
\textcolor{blue}{\href{https://doi.org/10.1103/PhysRevD.99.114003} {\textit{Phys. Rev. D} \textbf{99} (2019) 114003}}
[\textcolor{blue}{\href{https://arxiv.org/abs/1903.07115}{arXiv:1903.07115}}] 
[\textcolor{blue}{\href{https://inspirehep.net/literature/1725469}{\textsc{inSPIRE}}}].

\bibitem{Qian:2021neg} 
R.~Q.~Qian, Q.~Huang and X.~Liu,
\textit{Predicted $\Lambda\bar\Lambda$ and $\Xi^-\bar\Xi^{+}$ decay modes of the charmoniumlike $Y(4230)$},
\textcolor{blue}{\href{https://doi.org/10.1016/j.physletb.2022.137292} {\textit{Phys. Lett. B} \textbf{833} (2022) 137292}}
[\textcolor{blue}{\href{https://arxiv.org/abs/2111.13821}{arXiv:2111.13821}}] 
[\textcolor{blue}{\href{https://inspirehep.net/literature/1978833}{\textsc{inSPIRE}}}].

\bibitem{Ablikim:2013pgf} 
BESIII Collaboration,
\textit{Search for baryonic decays of $\psi(3770)$ and $\psi(4040)$},
\textcolor{blue}{\href{https://doi.org/10.1103/PhysRevD.87.112011} {\textit{Phys. Rev. D} \textbf{87} (2013) 112011}}
[\textcolor{blue}{\href{https://arxiv.org/abs/1305.1782}{arXiv:1305.1782}}] 
[\textcolor{blue}{\href{https://inspirehep.net/literature/1232386}{\textsc{inSPIRE}}}].

\bibitem{BESIII:2017kqg}
BESIII Collaboration,
\textit{Precision measurement of the $e^{+}e^{-}~\rightarrow~\Lambda_{c}^{+} \bar{\Lambda}_{c}^{-}$ cross section near threshold},
\textcolor{blue}{\href{https://doi.org/10.1103/PhysRevLett.120.132001} {\textit{Phys. Rev. Lett.} \textbf{120} (2018) 132001}}
[\textcolor{blue}{\href{https://arxiv.org/abs/1710.00150}{arXiv:1710.00150}}] 
[\textcolor{blue}{\href{https://inspirehep.net/literature/1628093}{\textsc{inSPIRE}}}].

\bibitem{Ablikim:2019kkp} 
BESIII Collaboration,
\textit{Measurement of the cross section for $e^+e^-\to\Xi^-\bar{\Xi}^+$ and observation of an excited $\Xi$ baryon},
\textcolor{blue}{\href{https://doi.org/10.1103/PhysRevLett.124.032002} {\textit{Phys. Rev. Lett.} \textbf{124} (2020) 032002}}
[\textcolor{blue}{\href{https://arxiv.org/abs/1910.04921}{arXiv:1910.04921}}] 
[\textcolor{blue}{\href{https://inspirehep.net/literature/1758883}{\textsc{inSPIRE}}}].

\bibitem{Wang:2021lfq}
BESIII Collaboration,
\textit{Study of baryon pair production at BESIII},
\textcolor{blue}{\href{https://doi.org/10.22323/1.385.0026} {\textit{PoS} \textbf{CHARM2020} 026 (2021)}} 
[\textcolor{blue}{\href{https://inspirehep.net/literature/1926588}{\textsc{inSPIRE}}}].


\bibitem{BESIII:2021ccp} 
BESIII Collaboration,
\textit{Measurement of the cross section for $e^+e^-\to\Lambda\bar{\Lambda}$ and evidence of the decay $\psi(3770)\to\Lambda\bar{\Lambda}$},
\textcolor{blue}{\href{https://doi.org/10.1103/PhysRevD.104.L091104} {\textit{Phys. Rev. D} \textbf{104} (2021) L091104}}
[\textcolor{blue}{\href{https://arxiv.org/abs/2108.02410}{arXiv:2108.02410}}] 
[\textcolor{blue}{\href{https://inspirehep.net/literature/1900124}{\textsc{inSPIRE}}}].


\bibitem{Wang:2022zyc}
X.~Wang and G.~Huang,
\textit{Electromagnetic form factor of doubly-strange hyperon},
\textcolor{blue}{\href{https://doi.org/10.3390/sym14010065} {\textit{Symmetry} \textbf{14} (2022) 65}} 
[\textcolor{blue}{\href{https://inspirehep.net/literature/2037456}{\textsc{inSPIRE}}}].


\bibitem{BESIII:2022kzc}
BESIII Collaboration,
\textit{Study of $e^+e^-\rightarrow\Omega^{-}\bar\Omega^{+}$ at center-of-mass energies from 3.49 to 3.67 GeV},
\textcolor{blue}{\href{https://doi.org/10.1103/PhysRevD.107.052003} {\textit{Phys. Rev. D} \textbf{107} (2023) 052003}}
[\textcolor{blue}{\href{https://arxiv.org/abs/2212.03693}{arXiv:2212.03693}}] 
[\textcolor{blue}{\href{https://inspirehep.net/literature/2611486}{\textsc{inSPIRE}}}].


\bibitem{Wang:2022bzl}
BESIII Collaboration,
\textit{Measurement of energy-dependent pair-production cross section and electromagnetic form Factors of a charmed baryon}, 
\textcolor{blue}{\href{https://doi.org/10.1103/PhysRevLett.131.191901} {\textit{Phys. Rev. Lett.} \textbf{131} (2023) 191901}} 
[\textcolor{blue}{\href{https://arxiv.org/abs/2307.07316}{arXiv:2307.07316}}] 
[\textcolor{blue}{\href{https://inspirehep.net/literature/2677290}{\textsc{inSPIRE}}}].


\bibitem{BESIII:2023rse}
BESIII Collaboration,
\textit{Measurement of the cross section of $e^+e^-\rightarrow\Xi^{-}\bar\Xi^{+}$ at center-of-mass energies between 3.510 and 4.843 GeV},
\textcolor{blue}{\href{https://doi.org/10.1007/JHEP11(2023)228}{\textit{JHEP} \textbf{11} (2023) 228}}
[\textcolor{blue}{\href{https://arxiv.org/abs/2309.04215}{arXiv:2309.04215}}] 
[\textcolor{blue}{\href{https://inspirehep.net/literature/2695411}{\textsc{inSPIRE}}}].


\bibitem{BESIII:2024umc}
BESIII Collaboration,
\textit{Measurement of Born cross section of $ {e}^{+}{e}^{-}\to {\Sigma}^{+}{\overline{\Sigma}}^{-} $ at center-of-mass energies between 3.510 and 4.951 GeV},
\textcolor{blue}{\href{https://doi.org/10.1007/JHEP05(2024)022}{\textit{JHEP} \textbf{05} (2024) 022}}
[\textcolor{blue}{\href{https://arxiv.org/abs/2401.09468}{arXiv:2401.09468}}] 
[\textcolor{blue}{\href{https://inspirehep.net/literature/2748736}{\textsc{inSPIRE}}}].


\bibitem{BESIII:kxls}
BESIII Collaboration,
\textit{Measurement of the cross sections of ${e}^{+}{e}^{-}\to {K}^{-}{\overline{\Xi}}^{+}\Lambda /{\Sigma}^0$ at center-of-mass energies between 3.510 and 4.914 GeV},
\textcolor{blue}{\href{https://doi.org/10.1007/JHEP07(2024)258} {\textit{JHEP} \textbf{07} (2024) 258}}
[\textcolor{blue}{\href{https://arxiv.org/abs/2406.18183}{arXiv:2406.18183}}] 
[\textcolor{blue}{\href{https://inspirehep.net/literature/2802333}{\textsc{inSPIRE}}}].

\bibitem{BESIII:2024sigma0}
BESIII Collaboration,
\textit{Measurement of Born cross section of ${e}^{+}{e}^{\ensuremath{-}}\ensuremath{\rightarrow}{\mathrm{\ensuremath{\Sigma}}}^{0}{\overline{\mathrm{\ensuremath{\Sigma}}}}^{0}$ at $\sqrt{s}=3.50--4.95$ GeV},
\textcolor{blue}{\href{https://doi.org/10.1103/PhysRevD.111.L051502}{\textit{Phys. Rev. D} \textbf{111} (2025) L051502}}
[\textcolor{blue}{\href{https://arxiv.org/abs/2412.20305}{arXiv:2412.20305}}] 
[\textcolor{blue}{\href{https://inspirehep.net/literature/2863767}{\textsc{inSPIRE}}}].


\bibitem{BESIII:2025hl}
BESIII Collaboration,
\textit{Measurement of the Born cross section for $ {e}^{+}{e}^{-}\to p{K}^{-}{K}^{-}{\overline{\Xi}}^{+} $at $ \sqrt{s}=3.5\hbox{--} 4.9 $ GeV},
\textcolor{blue}{\href{https://doi.org/10.1007/JHEP11(2025)111}{\textit{JHEP} \textbf{11} (2025) 111}}
[\textcolor{blue}{\href{https://arxiv.org/abs/2508.11276}{arXiv:2508.11276}}] 
[\textcolor{blue}{\href{https://inspirehep.net/literature/2961001}{\textsc{inSPIRE}}}].


\bibitem{BESIII:2025ruoyu}
BESIII Collaboration,
\textit{Search for charmonium(-like) states decaying into the $\Omega^-\bar{\Omega}^+$ final states},
\textcolor{blue}{\href{https://doi.org/10.1016/j.nuclphysb.2026.117388}{\textit{Nucl. Phys. B} \textbf{1025} (2026) 117388}}
[\textcolor{blue}{\href{https://arxiv.org/abs/2508.03454}{arXiv:2508.03454}}] 
[\textcolor{blue}{\href{https://inspirehep.net/literature/2957692}{\textsc{inSPIRE}}}].

\bibitem{zhang:2026ssb}
BESIII Collaboration,
\textit{Measurement of Born cross sections for $e^+e^-\to \Sigma^- \bar{\Sigma}^+$ at $\sqrt{s}\ =\ 3.51-4.95$ GeV and
observation of $\psi(3770)\to \Sigma^- \bar{\Sigma}^+$}, [\textcolor{blue}{\href{https://arxiv.org/abs/2602.23835}{arXiv:2602.23835}}].

\bibitem{ene1}
BESIII Collaboration,
\textit{Precision measurement of the integrated luminosity of the data taken by BESIII at center of mass energies between 3.810 GeV and 4.600 GeV},
\textcolor{blue}{\href{https://doi.org/10.1088/1674-1137/39/9/093001} {\textit{Chin. Phys. C} \textbf{39} (2015) 093001}}
[\textcolor{blue}{\href{https://arxiv.org/abs/1503.03408}{arXiv:1503.03408}}] 
[\textcolor{blue}{\href{https://inspirehep.net/literature/1351765}{\textsc{inSPIRE}}}].

\bibitem{BESIII:2022dxl}
BESIII Collaboration,
\textit{Measurement of integrated luminosities at BESIII for data samples at center-of-mass energies between 4.0 and 4.6 GeV},
\textcolor{blue}{\href{https://doi.org/10.1088/1674-1137/ac80b4} {\textit{Chin. Phys. C} \textbf{46} (2022) 113002}}
[\textcolor{blue}{\href{https://arxiv.org/abs/2203.03133}{arXiv:2203.03133}}] 
[\textcolor{blue}{\href{https://inspirehep.net/literature/2175754}{\textsc{inSPIRE}}}].

\bibitem{ene3}
BESIII Collaboration,
\textit{Luminosities and energies of $e^+ e^-$ collision data taken between 4.61 GeV and 4.95 GeV at BESIII},
\textcolor{blue}{\href{https://doi.org/10.1088/1674-1137/ac84cc} {\textit{Chin. Phys. C} \textbf{46} (2022) 113003}}
[\textcolor{blue}{\href{https://arxiv.org/abs/2205.04809}{arXiv:2205.04809}}] 
[\textcolor{blue}{\href{https://inspirehep.net/literature/2079606}{\textsc{inSPIRE}}}].


\bibitem{ene5}
BESIII Collaboration,
\textit{Determination of the number of $\psi(3686)$ events taken at BESIII},
\textcolor{blue}{\href{https://doi.org/10.1088/1674-1137/ad595b} {\textit{Chin. Phys. C} \textbf{48} (2024) 093001}}
[\textcolor{blue}{\href{https://arxiv.org/abs/2403.06766}{arXiv:2403.06766}}] 
[\textcolor{blue}{\href{https://inspirehep.net/literature/1623019}{\textsc{inSPIRE}}}].


\bibitem{ene4}
BESIII Collaboration,
\textit{Measurement of integrated luminosity of data collected at 3.773 GeV by BESIII from 2021 to 2024},
\textcolor{blue}{\href{https://doi.org/10.1088/1674-1137/ad70a0} {\textit{Chin. Phys. C} \textbf{48} (2024) 123001}}
[\textcolor{blue}{\href{https://arxiv.org/abs/2406.05827}{arXiv:2406.05827}}] 
[\textcolor{blue}{\href{https://inspirehep.net/literature/2796559}{\textsc{inSPIRE}}}].



\bibitem{besiii} 
BESIII Collaboration,
\textit{Design and construction of the BESIII detector},
\textcolor{blue}{\href{https://doi.org/10.1016/j.nima.2009.12.050} {\textit{Nucl. Instrum. Meth. A} \textbf{614} (2010) 345-399}}
[\textcolor{blue}{\href{https://arxiv.org/abs/0911.4960}{arXiv:0911.4960}}] 
[\textcolor{blue}{\href{https://inspirehep.net/literature/838149}{\textsc{inSPIRE}}}].

\bibitem{BEPCII} 
C. Yu \textit{et al.},
\textit{BEPCII performance and beam dynamics studies on luminosity},
\textcolor{blue}{\href{https://doi.org/10.18429/JACoW-IPAC2016-TUYA01} {\textit{Proceedings, 7th International Particle Accelerator Conference (IPAC 2016)} May 8-13 2016}} 
[\textcolor{blue}{\href{https://inspirehep.net/literature/1469857}{\textsc{inSPIRE}}}].

\bibitem{Ablikim:2019hff} 
BESIII Collaboration,
\textit{Future physics programme of BESIII},
\textcolor{blue}{\href{https://doi.org/10.1088/1674-1137/44/4/040001}{\textit{Chin. Phys. C} \textbf{44} (2020) 040001}}
[\textcolor{blue}{\href{https://arxiv.org/abs/1912.05983}{arXiv:1912.05983}}] 
[\textcolor{blue}{\href{https://inspirehep.net/literature/1770442}{\textsc{inSPIRE}}}].

\bibitem{EcmsMea}
J.~Lu, Y.~Xiao, and X.~Ji,
\textit{Online monitoring of the center-of-mass energy from real data at BESIII},
\textcolor{blue}{\href{https://doi.org/10.1007/s41605-020-00188-8}{\textit{Radiat. Detect. Technol. Methods} \textbf{4} (2020) 337–344}}.

\bibitem{EventFilter}
J.~W.~Zhang, L.~H.~Wu, and S.~S.~Sun \textit{et al.},
\textit{Suppression of top-up injection backgrounds with offline event filter in the BESIII experiment},
\textcolor{blue}{\href{https://doi.org/10.1007/s41605-022-00331-7}{\textit{Radiat. Detect. Technol. Methods} {\textbf 6} (2022) 289–293}} 
[\textcolor{blue}{\href{https://inspirehep.net/literature/2145899}{\textsc{inSPIRE}}}].


\bibitem{etof3}
 P.~Cao \textit{et al.},
\textit{Design and construction of the new BESIII endcap Time-of-Flight system with MRPC Technology},
\textcolor{blue}{\href{https://doi.org/10.1016/j.nima.2019.163053} {\textit{Nucl. Instrum. Meth. A} \textbf{953} (2020) 163053 }}
[\textcolor{blue}{\href{https://inspirehep.net/literature/1775466}{\textsc{inSPIRE}}}].


\bibitem{GEANT4} 
GEANT4 Collaboration,
\textit{GEANT4$-$a simulation toolkit},
\textcolor{blue}{\href{https://doi.org/10.1016/S0168-9002(03)01368-8}{\textit{Nucl. Instrum. Meth. A} \textbf {506} (2003) 250}}.

\bibitem{Huang:2022wuo}
K.~X.~Huang, Z.~J.~Li, Z.~Qian, J.~Zhu, H.~Y.~Li, Y.~M.~Zhang, S.~S.~Sun and Z.~Y.~You,
\textit{Method for detector description transformation to unity and application in BESIII},
\textcolor{blue}{\href{https://doi.org/10.1007/s41365-022-01133-8} {\textit{Nucl. Sci. Tech.} \textbf{33} (2022) 142}}
[\textcolor{blue}{\href{https://arxiv.org/abs/2206.10117}{arXiv:2206.10117}}] 
[\textcolor{blue}{\href{https://inspirehep.net/literature/2098691}{\textsc{inSPIRE}}}].

\bibitem{KKMC} 
S. Jadach, B. F. L. Ward and Z. Was,
\textit{Coherent exclusive exponentiation for precision Monte Carlo calculations},
\textcolor{blue}{ \href{https://doi.org/10.1103/PhysRevD.63.113009} {\textit{Phys. Rev. D} \textbf{63} (2001) 113009}}
[\textcolor{blue}{\href{https://arxiv.org/abs/hep-ph/0006359}{arXiv:hep-ph/0006359}}] 
[\textcolor{blue}{\href{https://inspirehep.net/literature/529540}{\textsc{inSPIRE}}}].

\bibitem{EVTGEN}
D. J. Lange,
\textit{The EvtGen particle decay simulation package},
\textcolor{blue}{\href{https://doi.org/10.1016/S0168-9002(01)00089-4}{\textit{Nucl. Instrum. Meth. A} \textbf{462} (2001) 152}}.

\bibitem{evtgen2} 
R. G. Ping,
\textit{Event generators at BESIII},
\textcolor{blue}{ \href{https://doi.org/10.1088/1674-1137/32/8/001} {\textit{Chin. Phys. C} \textbf{32} (2008) 599}}.

\bbt{PDG2020} 
Particle Data Group,
\textit{Review of particle physics},
\textcolor{blue}{\href{https://doi.org/10.1142/S0217751X26300115} {\textit{Int. J. Mod. Phys. A} \textbf{41} (2026) 2630011}} 
[\textcolor{blue}{\href{https://inspirehep.net/literature/3193100}{\textsc{inSPIRE}}}].

\bibitem{BESIII:2016nix}
BESIII Collaboration,
\textit{Study of $J/\psi$ and $\psi(3686)\rightarrow\Sigma(1385)^{0}\bar\Sigma(1385)^{0}$ and $\Xi^0\bar\Xi^{0}$},
\textcolor{blue}{\href{https://doi.org/10.1016/j.physletb.2017.04.048}{\textit{Phys. Lett. B} \textbf{770} (2017) 217}}
[\textcolor{blue}{\href{https://arxiv.org/abs/1612.08664}{arXiv:1612.08664}}] 
[\textcolor{blue}{\href{https://inspirehep.net/literature/1506414}{\textsc{inSPIRE}}}].

\bibitem{BESIII:2019dve}
BESIII Collaboration,
\textit{Observation of $\psi(3686)\rightarrow\Xi(1530)^{-}\bar\Xi(1530)^{+}$ and $\Xi(1530)^{-}\bar\Xi^{+}$},
\textcolor{blue}{\href{https://doi.org/10.1103/PhysRevD.100.051101} {\textit{Phys. Rev. D} \textbf{100} (2019) 051101}}
[\textcolor{blue}{\href{https://arxiv.org/abs/1907.13041}{arXiv:1907.13041}}] 
[\textcolor{blue}{\href{https://inspirehep.net/literature/1747092}{\textsc{inSPIRE}}}].

\bibitem{BESIII:2021aer} 
BESIII Collaboration,
\textit{Measurement of cross section for $e^+e^-\to\Xi^0\bar{\Xi}^0$ near threshold},
\textcolor{blue}{ \href{https://doi.org/10.1016/j.physletb.2021.136557} {\textit{Phys. Lett. B} \textbf{820} (2021) 136557}}
[\textcolor{blue}{\href{https://arxiv.org/abs/2105.14657}{arXiv:2105.14657}}] 
[\textcolor{blue}{\href{https://inspirehep.net/literature/1866233}{\textsc{inSPIRE}}}].

\bibitem{vtxfit} 
M. Xu \textit{et al.},
\textit{Decay vertex reconstruction and 3-dimensional lifetime determination at BESIII},
\textcolor{blue}{\href{https://doi.org/10.1088/1674-1137/33/6/005} {\textit{Chin. Phys. C} \textbf{33} (2009) 428}} 
[\textcolor{blue}{\href{https://inspirehep.net/literature/1122428}{\textsc{inSPIRE}}}].




\bibitem{Zhu:2008ca}
Y.~S.~Zhu,
\textit{Bayesian credible interval construction for Poisson statistics},
\textcolor{blue}{\href{https://doi.org/10.1088/1674-1137/32/5/007} {\textit{Chin. Phys. C} \textbf{32} (2008) 363}}
[\textcolor{blue}{\href{https://arxiv.org/abs/0812.2705}{arXiv:0812.2705}}] 
[\textcolor{blue}{\href{https://inspirehep.net/literature/805362}{\textsc{inSPIRE}}}].

\bibitem{Jegerlehner:2011ti}
F.~Jegerlehner and R.~Szafron,
\textit{$\rho^0-\gamma$ mixing in the neutral channel pion form factor $F_{\pi}^{e}$ and its role in comparing $e^+ e^-$ with $\tau$ spectral functions},
\textcolor{blue}{\href{https://doi.org/10.1140/epjc/s10052-011-1632-3}{\textit{Eur. Phys. J. C} \textbf{71} (2011) 1632}}
[\textcolor{blue}{\href{https://arxiv.org/abs/1101.2872}{arXiv:1101.2872}}] 
[\textcolor{blue}{\href{https://inspirehep.net/literature/884306}{\textsc{inSPIRE}}}].

 \bibitem{Kuraev:1985hb}
E.~A.~Kuraev and V.~S.~Fadin,
\textit{On radiative corrections to $e^+e^-$ single photon annihilation at high-energy},
\textcolor{blue}{ {\textit{Sov. J. Nucl. Phys.} \textbf{41} (1985) 466-472}} 
[\textcolor{blue}{\href{https://inspirehep.net/literature/217313}{\textsc{inSPIRE}}}].



\bibitem{Sun:2020ehv}
W.~Sun, T.~Liu, M.~Jing, L.~Wang, B.~Zhong, and W.~Song,
\textit{An iterative weighting method to apply ISR correction to $e^+ e^-$ hadronic cross section measurements},
\textcolor{blue}{\href{https://doi.org/10.1007/s11467-021-1085-6}{\textit{Front. Phys. (Beijing)} \textbf{16} (2021) 64501}}
[\textcolor{blue}{\href{https://arxiv.org/abs/2011.07889}{arXiv:2011.07889}}] 
[\textcolor{blue}{\href{https://inspirehep.net/literature/1830422}{\textsc{inSPIRE}}}].



\bibitem{BESIII:2021kwf}
BESIII Collaboration,
\textit{Measurement of the inclusive branching fraction for $\psi(3686)\rightarrow K_{S}^{0} + \text{anything}$},
\textcolor{blue}{\href{https://doi.org/10.1016/j.physletb.2021.136576} {\textit{Phys. Lett. B} \textbf{820} (2021) 136576}}
[\textcolor{blue}{\href{https://arxiv.org/abs/2106.08766}{arXiv:2106.08766}}] 
[\textcolor{blue}{\href{https://inspirehep.net/literature/1868813}{\textsc{inSPIRE}}}].

\bibitem{BESIII:2024xixi}
BESIII Collaboration,
\textit{Measurement of Born cross sections of $e^+e^- \to \Xi^0 \bar{\Xi}^0$ and search for charmonium (-like) states at $\sqrt{s}\ =\ 3.51-4.95{\rm GeV}$},
\textcolor{blue}{\href{https://doi.org/10.1007/JHEP11(2024)062} {\textit{JHEP} \textbf{11} (2024) 062}}
[\textcolor{blue}{\href{https://arxiv.org/abs/2409.00427}{arXiv:2409.00427}}] 
[\textcolor{blue}{\href{https://inspirehep.net/literature/2824143}{\textsc{inSPIRE}}}].


\bibitem{besiii:y4710}
BESIII Collaboration,
\textit{Observation of the $Y(4230)$ and evidence for a new vector charmoniumlike state $Y(4710)$ in ${e}^{+}{e}^{\ensuremath{-}}\ensuremath{\rightarrow}{K}_{S}^{0}{K}_{S}^{0}J/\ensuremath{\psi}$}
\textcolor{blue}{\href{https://link.aps.org/doi/10.1103/PhysRevD.107.092005} {\textit{Phys. Rev. D} \textbf{107} (2023) 092005}}
[\textcolor{blue}{\href{https://arxiv.org/abs/2211.08561}{arXiv:2211.08561}}] 
[\textcolor{blue}{\href{https://inspirehep.net/literature/2182758}{\textsc{inSPIRE}}}].


\end{thebibliography}
\end{document}